\documentclass[journal,twoside,web]{ieeecolor}
\usepackage{generic}
\usepackage{cite}
\usepackage{amsmath,amssymb,amsfonts}
\usepackage{algorithm}
\usepackage{algpseudocode}
\usepackage{graphicx}
\usepackage{textcomp}

\usepackage{makecell}
\usepackage{float}
\usepackage{subfig}
\usepackage[dvipsnames]{xcolor}
\definecolor{modify}{rgb}{1,0,0}
\definecolor{add}{rgb}{0, 0.3, 0}
\definecolor{transform}{rgb}{0,0,1}
\graphicspath{ {./pic/} }

\def\BibTeX{{\rm B\kern-.05em{\sc i\kern-.025em b}\kern-.08em
    T\kern-.1667em\lower.7ex\hbox{E}\kern-.125emX}}
\begin{document}
\title{Equilibrium Unit Based Localized Affine Formation Maneuver for Multi-agent Systems}
\author{Cheng~Zhu, 
	Xiaotao~Zhou,
	and Bing~Huang
	\thanks{This work was supported by the National Natural Science Foundation of China under Grant 52301370 and Grant 52371358.  (Cheng Zhu and Xiaotao Zhou contributed equally to this work.) (Corresponding author: Bing Huang.)}
	\thanks{Cheng Zhu and Bing Huang are with the School of Marine Science and Technology, Northwestern Polytechnical University, Xi’an 710072, China (e-mail: oliver\_rlz@hrbeu.edu.cn; binhuang@mail.nwpu.edu.cn).}
	\thanks{Xiaotao Zhou is with the National Key Laboratory of Autonomous Marine Vehicle Technology, Harbin Engineering University, Harbin 150001, China (e-mail: xtzhou@hrbeu.edu.cn.)}%
}

\maketitle

\begin{abstract}
Current affine formation maneuver of multi-agent systems (MASs) relies on the affine localizability determined by generic assumption for nominal configuration and global construction manner. This does not live up to practical constraints of robot swarms. In this paper, an equilibrium unit based structure is proposed to achieve affine localizability. In an equilibrium unit, existence of non-zero weights between nodes is guaranteed and their summation is proved to be non-zero. To remove the generic assumption, a notion of layerable directed graph is introduced, based on which a sufficient condition associated equilibrium unit is presented to establish affine localizability condition. Within this framework, distributed local construction manner is performed by a designed equilibrium unit construction (EUC) method. With the help of localized communication criterion (LCC) and localized sensing based affine formation maneuver control (LSAFMC) protocol, self-reconstruction capability is possessed by MASs when nodes are added to or removed from the swarms.
\end{abstract}


\section{Introduction}

Affine formation maneuver control concept is becoming prevailing for multi-agent systems (MASs). Its popularity can be attributed to the the following reasons: 1) Compared with formation keeping control methods \cite{ren2006consensus,cheng2018fully,pan2023improved}, it better aligns with the unexpected environments, where pop-up obstacles put forward the maneuverability of formation pattern at high demands. 2) Superior to others formation maneuver control methods \cite{aranda2015coordinate,zhao2015bearing,chen2020angle}, it possesses arbitrary configuration regulation, including translation, rotation and scaling. The core of achieving affine formation maneuver lies in endowing the affine localizability to MASs. Specifically, fundamental steps of designing affine localizability is summarized as follows:
\begin{enumerate}
	\item Establish flexible conditions for MASs' nominal framework to satisfy affine localizability. 
	\item Construct MASs' nominal framework by feasible methods to achieve affine localizability.
\end{enumerate}

The affine formation control problem is first proposed in \cite{lin2015necessary} under both directed graph and undirected graph, where conditions and control laws are given for converging MASs' trajectories to affine image of nominal configuration. Inspired by \cite{lin2015necessary}, \cite{zhao2018affine} and \cite{xu2020affine} extend the concept of affine formation to affine formation maneuver by addressing the leader selection problem and introduce the notion of affine localizability. Based on the concept of affine localizability, various researches have been investigated to enrich the application of affine formation maneuver \cite{zhu2022distributed,zhao2023specified,chen2020distributed,xiao2022framework,gao2022practical,li2020layered,wang2021affine,lin2021unified}. Nevertheless, all the existing results rely on the generic condition of nominal configuration, which dramatically constrains the selection range of candidate nominal framework. Specifically, as stated in \cite{zhao2018affine}, the generic condition is an over-condition for achieving affine localizability, i.e., there exist nominal framework with non-generic nominal configuration but still has affine localizability. On the one hand, in practical robotic systems, alternative desire configurations of MASs are belong to the feasible region determined by environment constraints and mission requirements. On the other hand, actual configurations of MASs stem from the affine image of the nominal configuration. Thus, the achievement of pre-assigned formation tasks depends on the matching degree between the feasible region and affine image defined above, once the affine image cannot reach the feasible region, the pre-specified maneuvering requirements will not be satisfied. Therefore, a more flexible condition for nominal configuration without generic assumption is preferred to endow MASs with affine localizability. 

From the perspective of construction method, all the existing methods rely on pre-assigned nominal configuration and edges, and then calculate the corresponding weight of each edge. To be specific, for a given nominal configuration and an incidence matrix the construction problem to achieve affine localizability is modeled as a semi-definite programming \cite{alfakih2011bar} and a linear matrix inequality problem \cite{zhao2018affine}. However, such method comes with two inherent limitations that constrain its practical application. Firstly, global construction manner is required to implement such method. Specifically, nominal configuration and edges are determined by central station in advance and then non-zero weights are broadcasted to each node after calculation. Thus, global assessment and centralized decision-making behaviors are utilized when such method is implemented. Note that in practical engineering, not all formation members are equipped with the interaction access to central station. Therefore, approaches that depend on global assessment may be failed to reach the real-world hardware demands, especially when they are applied to large-scale MASs. Secondly, such method does not possess self-reconstruction capability. To maintain the affine localizazbility, scenarios of adding and removing nodes from the existing nominal framework will resulting in global reconstruction of the nominal framework. It means that reconstruction information transmitted from central station will need several hops to reach the robots corresponding to the remaining nodes. As a result, some robots in the MASs have a slow reaction to unexpected situations, such as pop-up obstacles. Therefore, system stability may not be guaranteed under reconstruction manner that rely on global updating. Consequently, it is necessary to establish a feasible construction method and self-reconstruction method that with distributed local manner.

Motivated by the above observations, this paper aims to build new structure for achieving affine localizability. The focus is on the model of equilibrium unit, which provides possibility for condition establish and construction method from local perspective. The equilibrium unit is a point set that determined by the dependence properties: 1) Affine dependence of points. 2) Linear dependence of displacement between points. Benefiting from the designed equilibrium unit, existence of non-zero weights between nodes is proved. Furthermore, the summation of non-zero weights is naturally to be non-zero within equilibrium unit, which offers the cornerstone for achieving affine localizability. To overcome the first challenge, that is, establishing a more flexible condition without generic assumption, a special graph-theoretic property, called layerable directed graph, is introduced to help the accomplishment of the affine localizability. Under the layerable directed graph topology, a sufficient condition that only associates with equilibrium unit is presented. Under the designed equilibrium unit based affine localizability condition, an equilibrium unit construction (EUC) method is subsequently proposed to transform the global construction manner to distributed local manner. Specifically, each follower node only need to self-determine the in-neighbors and their corresponding non-zero weights by examining if the selected in-neighbors and itself consist of an equilibrium unit or not. Considering the requirement of self-reconstruction capability, a novel localized communication criterion (LCC) is developed in this paper to cope with scenarios of adding and removing nodes. Then, a robot-oriented localized sensing based affine formation maneuver control (LSAFMC) is proposed to uniformly solve the system stability problem both in scenarios of adding nodes to the existing nominal framework and removing nodes from the existing nominal framework.

The main contributions of this work are threefolds: 1) Local equilibrium unit is proposed to establish affine localizability condition. Generic assumption for nominal configuration is removed, and instead, layerable directed graph is introduced as premise. 2) Distributed local construction manner for affine localizability is realized by introducing the EUC. Global assessment of information is not necessary, resulting in less reliance on central station of robots. 3) Self-reconstruction capability is equipped during MASs' affine formation maneuver by developing LCC and LSAFMC protocol. Affine localizability and system stability are maintained in scenario of adding nodes to (removing nodes from) the existing nominal framework.

\section{From Generic Condition to Equilibrium Unit}
Given multi-agent systems (MASs) consisted of $N$ agents. The vertex set is denoted by $\mathcal{V}=\left\{ \mathcal{V}_1, \mathcal{V}_2, \ldots ,\mathcal{V}_N \right\}$. Suppose the first $N_{l}<N$ nodes are leaders and the rest $N_{f} = N - N_{l}$ are followers. Define the leader set as $\mathcal{V}_l=\left\{ \mathcal{V}_1, \mathcal{V}_2, \ldots, \mathcal{V}_{N_{l}}\right\}$ and the follower set as $\mathcal{V}_f=\mathcal{V}/\mathcal{V}_l$.

Let $\mathcal{G}=({\mathcal{V}}, \mathcal{E}, \mathcal{W})$ be a directed graph, where $\mathcal{E} \subseteq {\mathcal{V}} \times {\mathcal{V}}$ denotes an edge set. $\left( \mathcal{V}_i,\mathcal{V}_j \right) \in \mathcal{E}$ means that $\mathcal{V}_{j}$ can assess the relative state regarding $\mathcal{V}_{i}$. For $\mathcal{V}_i \in {\mathcal{V}}$, define $\mathcal{N}^{\text{in}}_{i}=\left\{ \mathcal{V}_j \in {\mathcal{V}}:\left( \mathcal{V}_j,\mathcal{V}_i \right)\in \mathcal{E} \right\}$ and $\mathcal{N}^{\text{out}}_{i}=\left\{ \mathcal{V}_j \in {\mathcal{V}}:\left( \mathcal{V}_i,\mathcal{V}_j \right)\in \mathcal{E} \right\}$ as the in-neighbor set out-neighbor of $\mathcal{V}_{i}$. $\mathcal{W}$ is the weight set of $\mathcal{G}$ with $( \mathcal{V}_j,\mathcal{V}_i ) \in {\mathcal{V}} \times {\mathcal{V}}$, $w_{i,j}$ defined as
\begin{equation}
	\label{affine 0}
	\begin{aligned} 
		\begin{cases}
			w_{i,j} \neq 0, \text{ if }\left( \mathcal{V}_j,\mathcal{V}_i \right) \in \mathcal{E}\\
			w_{i,j} = 0, \text{ if }\left( \mathcal{V}_j,\mathcal{V}_i \right) \notin \mathcal{E}
		\end{cases}.
	\end{aligned}
\end{equation}
By defining $W = \left[w_{i,j}\right]$, the Laplacian matrix associated with the directed graph $\mathcal{G}$ is defined as ${\Omega}$:
\begin{align} 
	{\Omega} = \text{Deg}(W) - W = \left[ \begin{array}{c|c}
		{\Omega}_{ll} & {\Omega}_{lf} \\ \hline 
		{\Omega}_{fl} & {\Omega}_{ff} 
	\end{array} \right],\nonumber
\end{align}
where $\Omega_{ll}\in\mathbb{R}^{N_{l}\times N_{l}}$ and $\Omega_{ff}\in\mathbb{R}^{{N_{f}\times N_{f}}}$.

We define the nominal framework as the directed graph $\mathcal{G}$ associated with the nominal configuration $\chi \in \mathbb{R}^{dN}$, i.e., $(\mathcal{G},\chi)$. Let $p^{*}(t) \in \mathbb{R}^{dN}$ be the desired configuration during, which is generate from the affine transformation of $\chi$. In this context, we have
\begin{equation} \label{affine 2}
	p^{*}(t) = ( I_{N}\otimes A(t) )\chi + 1_{N}\otimes b(t),
\end{equation}
where $A(t)\in \mathcal{C}^{1}\left(\mathbb{R}_{\ge 0}, \mathbb{R}^{d\times d}\right)$ and $b(t)\in \mathcal{C}^{1}\left(\mathbb{R}_{\ge 0}, \mathbb{R}^{d}\right)$ denote gains inducing configuration transformations. Correspondingly, one desired framework of $(\mathcal{G},\chi)$ is defined as $(\mathcal{G},p^{*}(t))$.

In the next section, we will recall some classical concepts concerning the so-called affine localizability of directed nominal framework as well as relative theorems proposed in \cite{zhao2018affine,xu2020affine,lin2015necessary}, and in the subsequent subsection, we will report a series of new results on directed nominal framework that attempt to achieve the affine localizability from the perspective of local construction.

\subsection{Affine Localizability under Generic Condition}
\noindent \textbf{Definition 1.} \textit{(Equilibrium weight)} \cite{zhao2018affine} For the nominal framework $(\mathcal{G},\chi)$ with $\mathcal{G}=({\mathcal{V}},\mathcal{E}, \mathcal{W})$, weight set $\mathcal{W}$ is called an equilibrium weight set if it satisfies:
\begin{align}
	\sum\limits_{j\in\mathcal{N}^{\text{in}}_{i}}\omega_{i,j}(\chi_{i}-\chi_{j}) = 0, i \in {\mathcal{V}}. \label{equilibrium weight set}
\end{align}

\noindent \textbf{Definition 2.} \textit{(Affine image)} \cite{lin2015necessary} The affine image of the nominal configurations $\chi$ is defined as:
\begin{align}
	\mathcal{A}(\chi) = \{({{I}_{N}} \otimes A)\chi+{{1}_{N}}\otimes b|\forall A\in \mathbb{R}^{2\times2},b\in {{\mathbb{R}}^{2}}\}. \label{affine image 1}
\end{align}
By defining the homogeneous configuration $\bar{\chi} = \text{col}\{\bar{\chi}_{j}\}_{j \in \mathcal{V}}$ with $\bar{\chi}_{j} = \text{col}\{\chi_{j},1\}$, we have
\begin{equation}
	\begin{aligned}
		\mathcal{A}(\chi) =  \text{Im}(P(\bar{\chi}) \otimes I_{d}),\text{
		} P(\bar{\chi}) = 
		\begin{bmatrix}
			\bar{\chi}_{1}^{\text{T}} \\
			\bar{\chi}_{2}^{\text{T}} \\
			\vdots \\
			\bar{\chi}_{N}^{\text{T}}
		\end{bmatrix}.
		\label{affine image 2}
	\end{aligned}
\end{equation}

Then, we present the concept of affine localizability as follows:

\noindent \textbf{Definition 3.} \textit{(Affine localizability)} \cite{zhao2018affine} The nominal framework has affine localizability if the following conditions are satisfied:
\begin{enumerate}
	\item For any $p = \text{col}\{p_{l},p_{f}\} \in \mathcal{A}(\chi)$, $p_{f}$ can be uniquely determined by $p_{l}$.
	\item The Laplacian matrix $\Omega$ satisfying \eqref{equilibrium weight set} such that $\Omega_{ff}$ is invertible.
\end{enumerate}

To achieve the affine localizability, \cite{xu2020affine} has proposed an affine localizability condition for directed nominal framework under the generic assumption:

\noindent \textbf{Lemma 1.} \cite{xu2020affine} Suppose the nominal configuration $\chi$ is generic with $\{\chi_{i}\}^{N_{l}}_{i=1}$ affinely span $\mathbb{R}^{d}$, then, the nominal framework $(\mathcal{G},\chi)$ has affine localizability if and only if $\mathcal{G}$ is $(d + 1)$-rooted.

\begin{figure}[!t]
	\centering
	\includegraphics[width=3.2in]{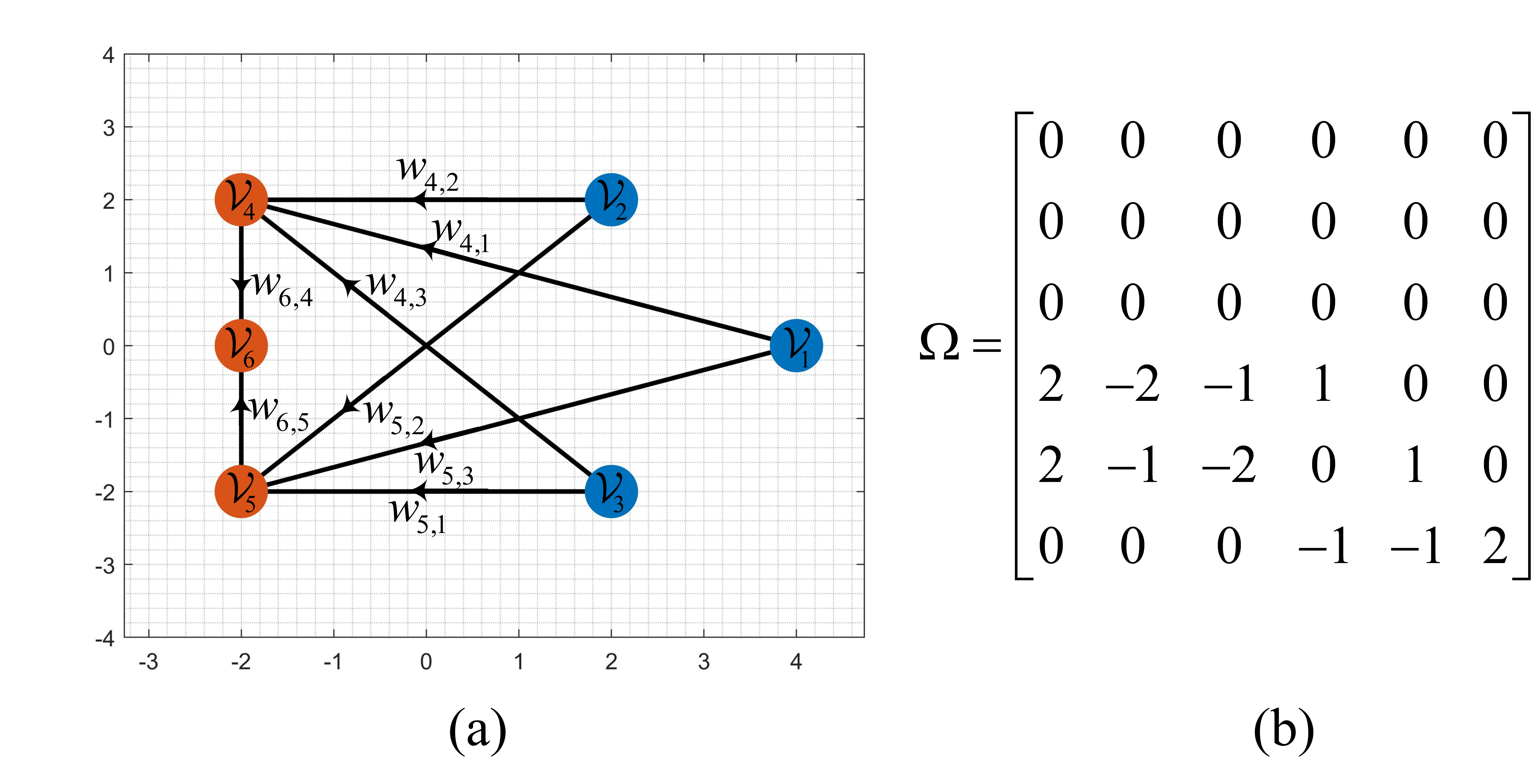}
	\caption{{Example to illustrate the nominal framework with affine localizability in $\mathbb{R}^{2}$. Specifically, $\mathcal{V}_{1}$, $\mathcal{V}_{2}$ and $\mathcal{V}_{3}$ are set as leaders and $\mathcal{V}_{4}$, $\mathcal{V}_{5}$ and $\mathcal{V}_{6}$ are set as followers. Positions of nodes are presented in (a), thus, we can conclude that leaders affinely span $\mathbb{R}^{2}$. Then, condition 1) in \textit{Definition 3} is satisfied. According to the associated Laplacian matrix presented in (b), it is easy to obtain that $\Omega_{ff}$ is invertible. Then, condition 2) in \textit{Definition 3} is satisfied. Note that the nominal configuration of the nominal framework is not generic because there exist collinear nodes, i.e., $\mathcal{V}_{4}$, $\mathcal{V}_{5}$ and $\mathcal{V}_{6}$ collineate in $\mathbb{R}^{2}$.}}
	\label{graph}
\end{figure}

\textit{Lemma 1} provides a equivalent condition for affine localizability on the premise of generic nominal configuration with leaders affinely span $\mathbb{R}^{d}$. Nevertheless, this method comes with two inherent limitations that constrain its applicability. The first limitation sterms from the over-condition of generic. As stated in \cite{zhao2018affine}, in undirected graph, the affine localizability may still be realized even the nominal configuration is not generic. Such problem also exists in directed graph. Fig. \ref{graph} shows a nominal framework with affine localizability but not in generic position. The second limitation pertains to the global construction method. Specifically, to achieve the affine localizability by \textit{Lemma 1}, we are requested to globally determine the nominal configuration in generic position and the graph with $(d + 1)$-rooted, which means that the actions of adding and removing a node will result in globally reconstruction of the nominal framework. Thus, there lacks concrete construction steps especially when considering the maintenance of affine localizability after adding and removing nodes in the existing nominal framework.

These observations inspire us to locally construct a nominal framework with affine localizability step by step. For addressing the first limitation, we attempt to extend the affine localizability to non-generic nominal configuration and weaken the $(d + 1)$-rooted condition. Considering the second limitation, our goal is to establish an reconfigurable method such that the affine localizability can be maintained when appropriately adding new nodes and removing old nodes. However, global construction method is no longer suitable for addressing these challenges, which motivates us to propose the concept of equilibrium unit in the next subsection.

\subsection{Affine Localizability under Equilibrium Unit}

\begin{figure}[!t]
	\centering
	\subfloat[]{\includegraphics[width=1.5in]{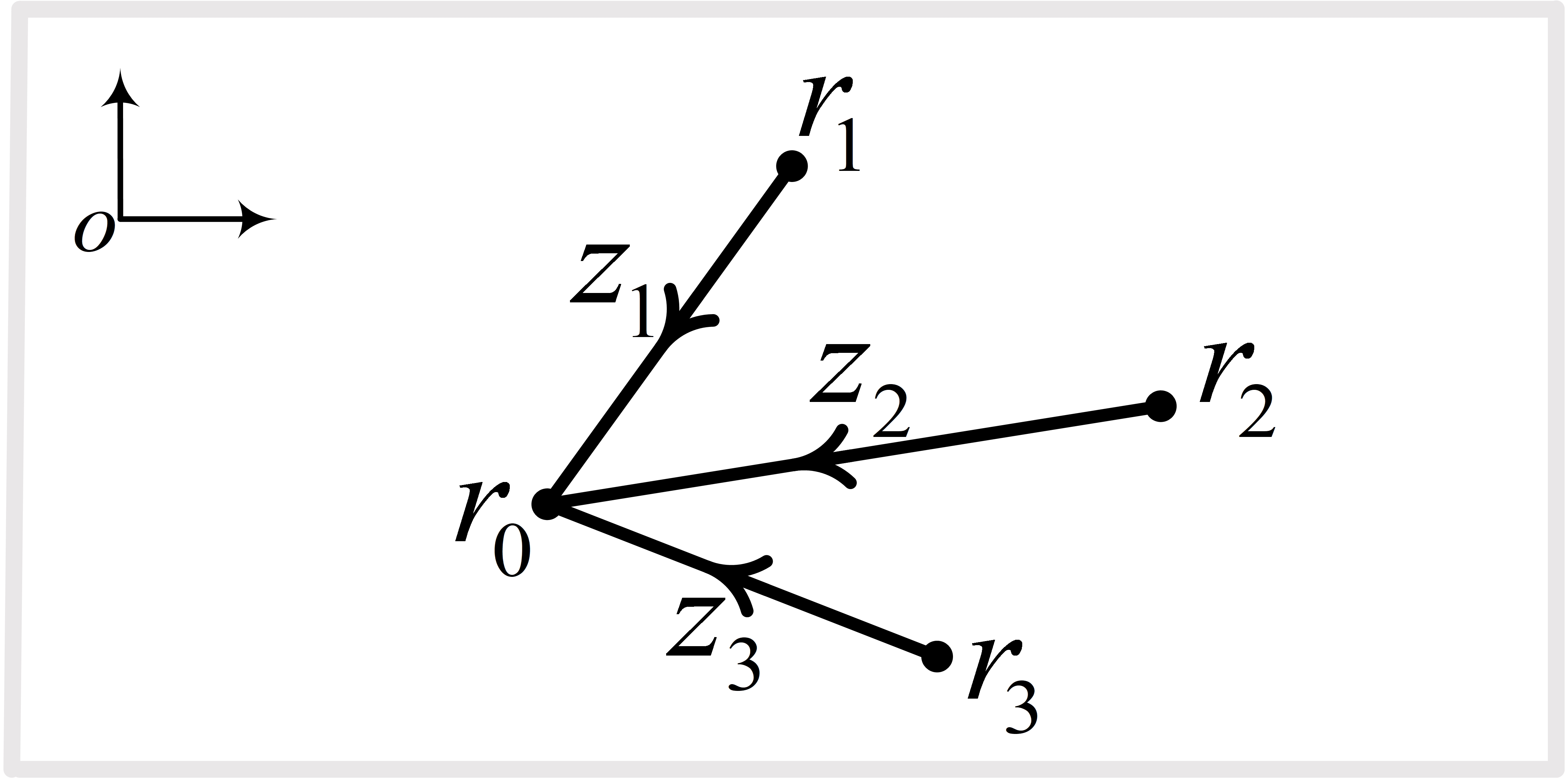}%
		\label{2_rep1}}
	\hfil
	\subfloat[]{\includegraphics[width=1.5in]{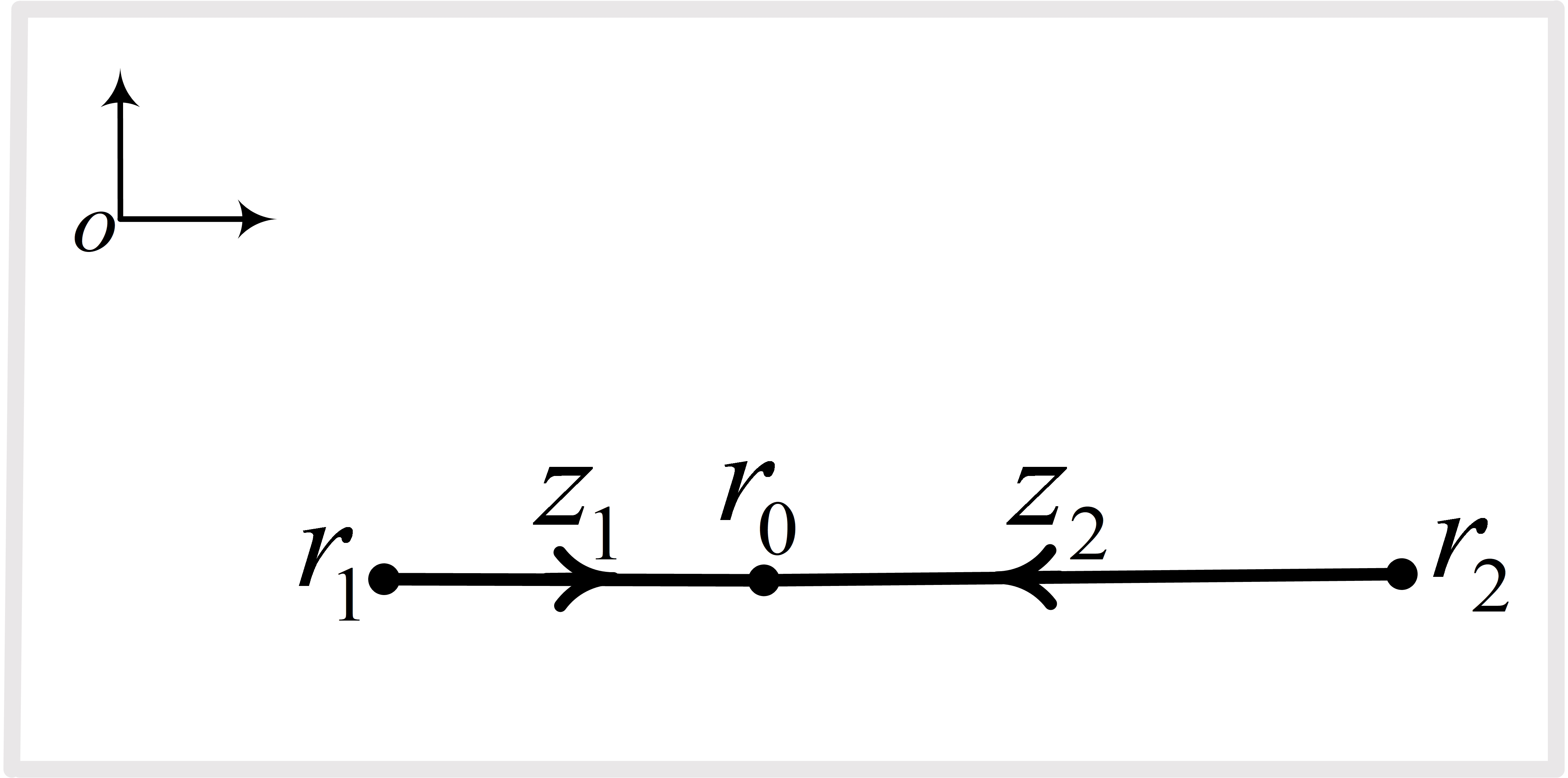}%
		\label{2_rep2}}
	\caption{(a) {Representation element in $\mathcal{U}^{3}(\mathbb{R}^{2})$. (b) Representation element in $\mathcal{U}^{2}(\mathbb{R}^{2})$.}}
	\label{pic1}
\end{figure}

\begin{figure}[!t]
	\centering
	\subfloat[]{\includegraphics[width=1.5in]{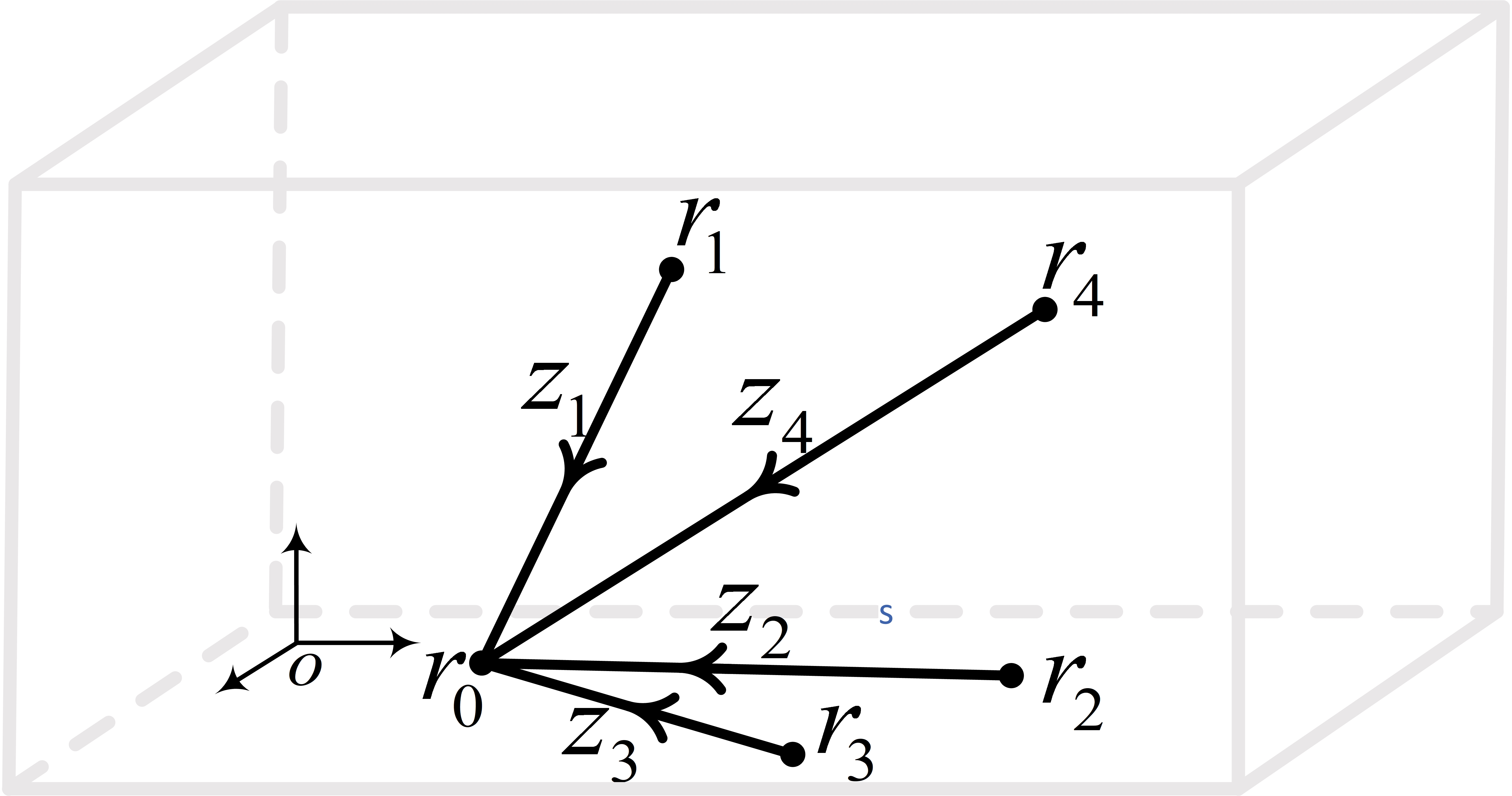}%
		\label{3_rep1}}
	\hfil
	\subfloat[]{\includegraphics[width=1.5in]{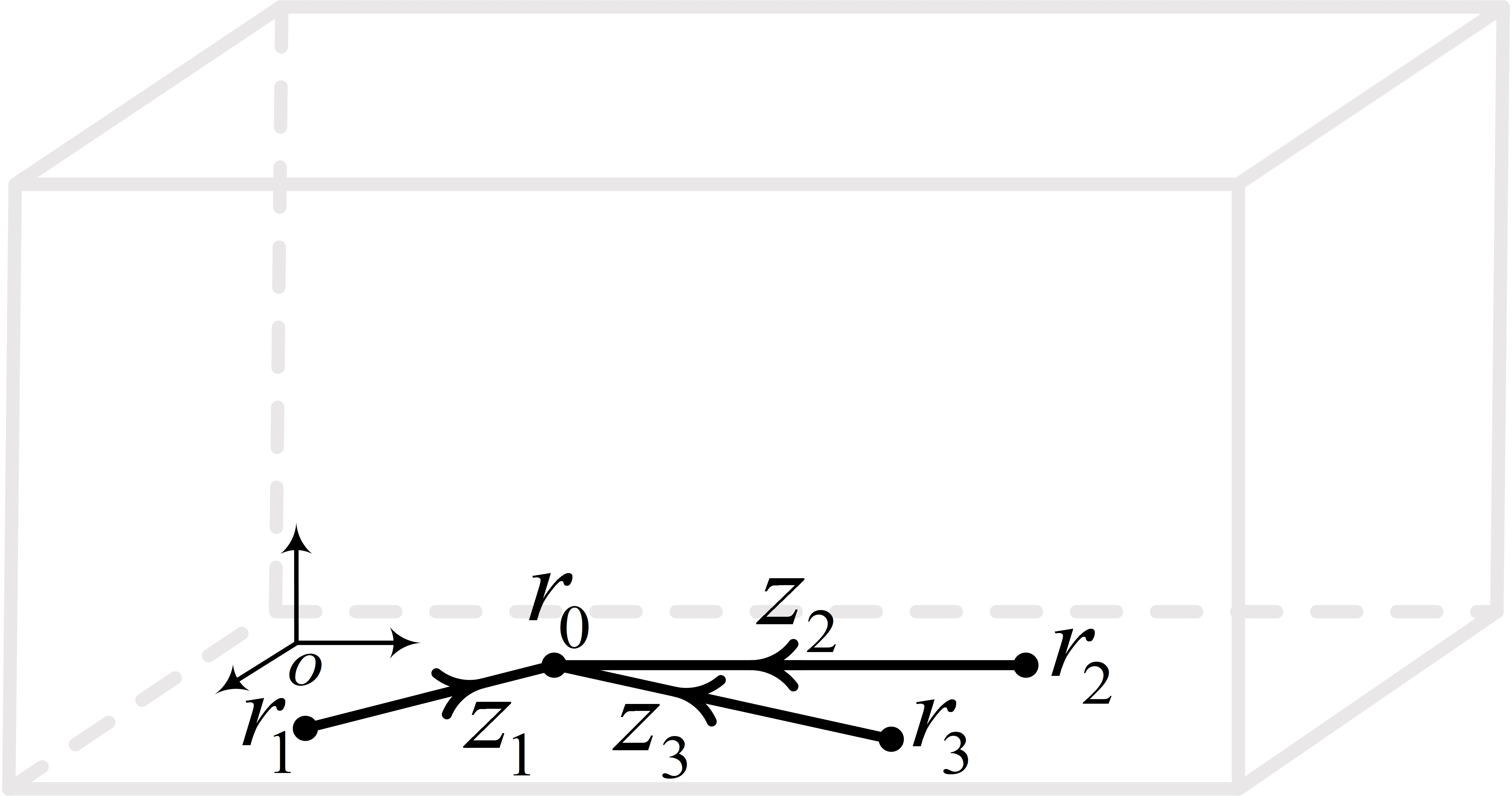}%
		\label{3_rep2}}
	\hfil
	\subfloat[]{\includegraphics[width=1.5in]{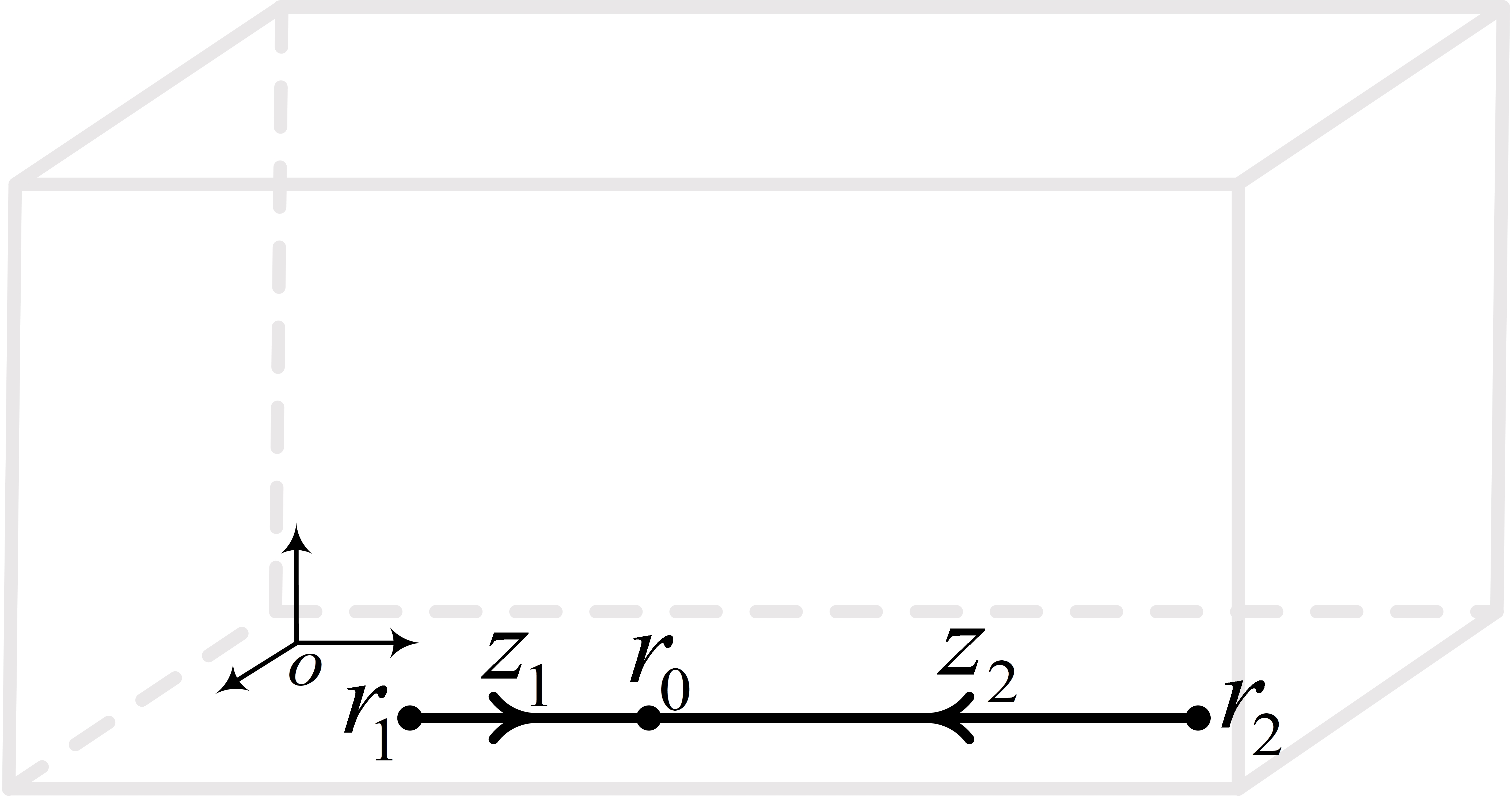}%
		\label{3_rep3}}
	\caption{(a) {Representation element in $\mathcal{U}^{4}(\mathbb{R}^{3})$. (b) Representation element in $\mathcal{U}^{3}(\mathbb{R}^{3})$. (c) Representation element in $\mathcal{U}^{2}(\mathbb{R}^{3})$.}}
	\label{pic2}
\end{figure}

When considering the possibility of local construction of nominal framework, an important aspect is how to choose suitable in-neighbors and locate the relative position between node and its in-neighbors such that the established overall nominal framework will possess the affine localizability. Thus, we firstly propose the following location principle between node and its in-neighbors, which is defined as equilibrium unit.

\noindent \textbf{Lemma 2.} \textit{($M_{d}$-equilibrium unit in $\mathbb{R}^{d}$)} For $R = \{r_{i}\in\mathbb{R}^{d}\}_{i = 0}^{M_{d}}$ with $2 \le M_{d} \le d + 1$, there exists $\{h_{j}\in\mathbb{R}_{\ne 0}\}_{j = 1}^{M_{d}}$ such that
\begin{align}
	&\sum\limits_{j = 1}^{M_{d}}h_{j}(r_{0}-r_{j}) = 0_{d},\label{1}\\
	&\sum\limits_{j = 1}^{M_{d}} h_{j} \ne 0, \label{2}
\end{align} 
if $R$ is a $M_{d}$-equilibrium unit in $\mathbb{R}^{d}$, i.e., the following conditions are satisfied:
\begin{enumerate}
	\item $\forall R_{j} = R / \{r_{j}\}$, $j = 0,1,\ldots,M_{d}$ is affinely independent.
	\item $Z = \{z_{i} = r_{0}-r_{i}\}_{i = 1}^{M_{d}}$ is linearly dependent.
\end{enumerate}
The set consisted of all $M_{d}$-equilibrium units in $\mathbb{R}^{d}$ is called the $M_{d}$-equilibrium unit set in $\mathbb{R}^{d}$ and is denoted by $\mathcal{U}^{M_{d}}(\mathbb{R}^{d})$. For $M_{d} = 2,3,\ldots,d+1$, $\bar{\mathcal{U}}(\mathbb{R}^{d}) = \cup_{M_{d} = 2}^{d+1}\mathcal{U}^{M_{d}}(\mathbb{R}^{d})$ is defined as the set of all equilibrium units in $\mathbb{R}^{d}$.

\noindent \textit{Proof.} Please refer to \textit{Appendix A}.

\textit{Remark 1.} By applying condition 1) in \textit{Lemma 2}, equilibrium unit candidates can be selected from the perspective of point distribution. Furthermore, within the allowable shapes formed by condition 1), we can subsequently determine the equilibrium unit by verifying condition 2) in \textit{Lemma 2}.

\textit{Remark 2.} When $M_{d} = d + 1$, conditions for $(d+1)$-equilibrium unit are equivalent to that $R$ is in general position in $\mathbb{R}^{d}$. When $M_{d} = 2$, we call $2$-equilibrium unit as the minimal equilibrium unit in $\mathbb{R}^{d}$.

\noindent \textbf{Corollary 1.} \textit{(2-dimension and 3-dimension cases)} When $d = 2$ and $d = 3$, representation elements in $\mathcal{U}^{M_{2}}(\mathbb{R}^{2})$ and $\mathcal{U}^{M_{3}}(\mathbb{R}^{3})$ are illustrated in Fig. \ref{pic1} and Fig. \ref{pic2}, respectively.

Based on the above discussion and \textit{Lemma 2}, we are able to give the main results of this subsection.

\noindent \textbf{Lemma 3.} \textit{(Layerable directed graph)} For any given directed graph $\mathcal{G}=({\mathcal{V}},\mathcal{E}, \mathcal{W})$, let $M$ be the number of nodes involved in the longest path in $\mathcal{G}$. Then, $\mathcal{G}$ is layerable, i.e., there exist unique classification of nodes $\mathcal{H}_{1},\mathcal{H}_{2},\ldots,\mathcal{H}_{M}$ such that
\begin{enumerate}
	\item $\cup_{i = 1}^{M}\mathcal{H}_{i} = \mathcal{V}$.
	\item $\mathcal{H}_{i} \cap \mathcal{H}_{j} = \emptyset$, for $\forall i \neq j$, $i,j = 1,2,\ldots,M$.
	\item For $\forall \mathcal{V}_{i} \in \mathcal{H}_{j}$, $j = 2,3,\ldots,M$, in-neighbors of $\mathcal{V}_{i}$ are all belong to $\cup_{k = 1}^{j - 1}\mathcal{H}_{k}$ and there is at least one in-neighbor belongs to $\mathcal{H}_{j - 1}$.
\end{enumerate}
if and only if there are no repeated nodes in any path in $\mathcal{G}$. $\mathcal{H}_{1},\mathcal{H}_{2},\ldots,\mathcal{H}_{M}$ are called layers of $\mathcal{G}$.

\noindent \textit{Proof.} Please refer to \textit{Appendix B}.

\noindent \textbf{Theorem 1.} \textit{(Affine localizability under equilibrium unit condition)} Suppose the nominal configuration $\chi$ satisfying $\{\chi_{i}\}_{i=1}^{N_{l}}$ affinely span $\mathbb{R}^{d}$. Then, the nominal framework $(\mathcal{G},\chi)$ has affine localizability if the directed graph $\mathcal{G}$ is layerable and each follower with its in-neighbors consist of an equilibrium unit in $\mathbb{R}^{d}$.

\noindent \textit{Proof.} Please refer to \textit{Appendix C}.

\noindent \textbf{Theorem 2.} \textit{(Affine localizability under equilibrium unit construction)} For the nominal framework $(\mathcal{G},\chi)$ with $\mathcal{G}=({\mathcal{V}},\mathcal{E}, \mathcal{W})$ and $N \ge d+2$, we can construct the nominal configuration and weight set such that $(\mathcal{G},\chi)$ has affine localizability by the following EUC steps:

\noindent \textit{Step 1.} Set the first $(d+1)$ nodes as leaders and $\{\chi_{i}\}_{i = 1}^{d+1}$ affinely span $\mathbb{R}^{d}$.

\noindent \textit{Step 2.} For $\mathcal{V}_{k}$, $k = d + 2, d + 3, \ldots, N$, merge one $M_{d}$-equilibrium unit $R = \{r_{i}\}_{i = 0}^{M_{d}}$ with $2 \le M_{d} \le d + 1$ from $\mathcal{U}^{M_{d}}(\mathbb{R}^{d})$ into the constructed part such that $\chi_{k}$ coincides with $r_{0}$ and $\{r_{i}\}_{i = 1}^{M_{d}}$ coincide with the existing points $\{\chi_{v_{i}}\}_{i = 1}^{M_{d}}$. Calculate the corresponding non-zero weights $\{w_{k,v_{j}}\}_{j=1}^{{M_{d}}}$ by \eqref{1} and \eqref{2}.

\noindent \textit{Step 3.} Repeat \textit{Step 2} until the last node.

\noindent \textit{Proof.} Please refer to \textit{Appendix D}.

\begin{figure}[!t]
	\centering
	\includegraphics[width=3in]{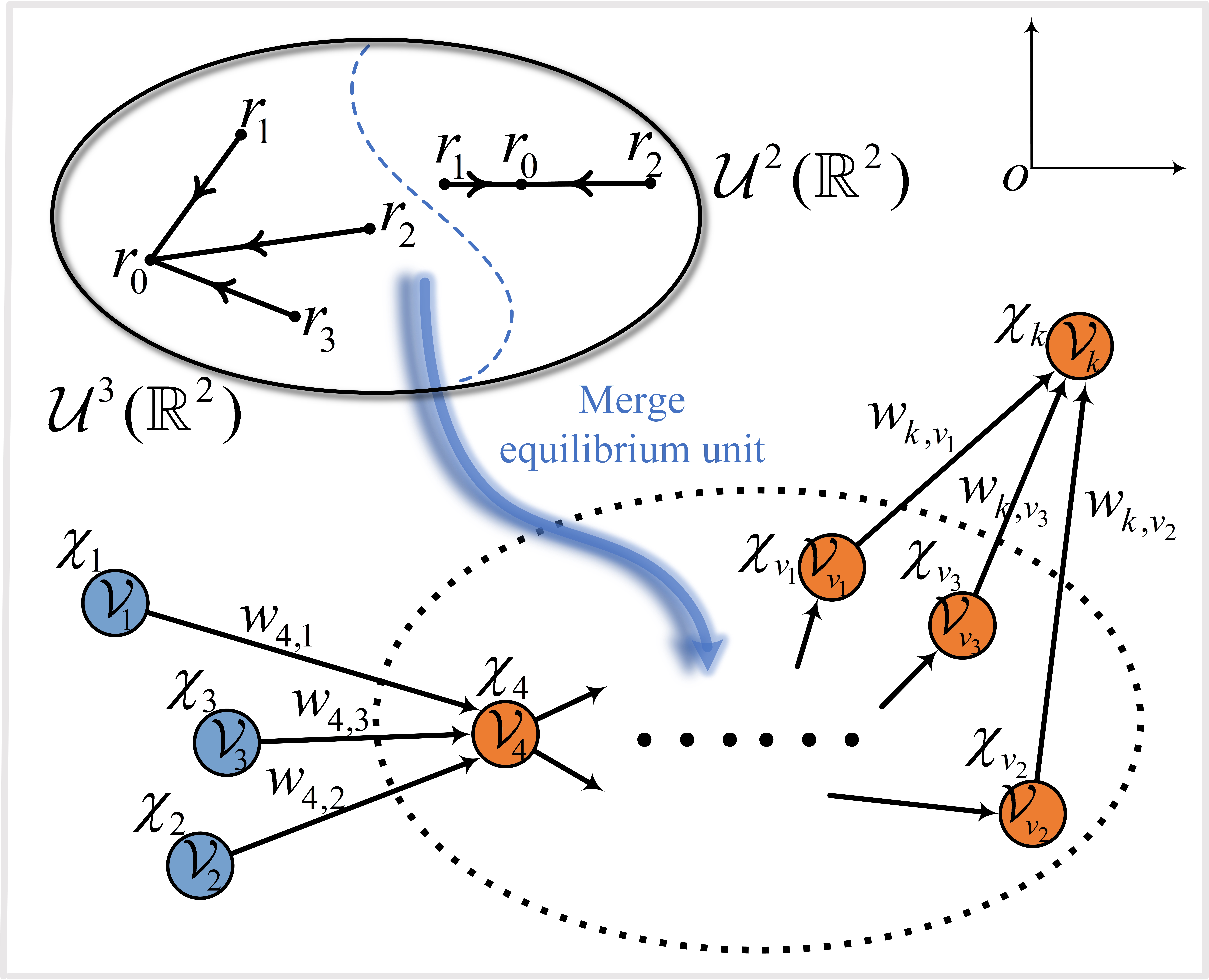}
	\caption{{Illustration of EUC procedure proposed in \textit{Theorem 1} ($\mathbb{R}^{2}$ case).}}
	\label{EUC}
\end{figure}

An example of the EUC procedure is presented in Fig. \ref{EUC}. An important feature of EUC is that each follower node has at least two in-neighbors and at most $(d + 1)$ in-neighbors.

\textit{Remark 3.} Our proposed construction method contributes the affine localizability with the following key properties:
\begin{enumerate}
	\item From perspective of affine localizability condition, generic condition for nominal configuration is removed. Compared with the existing affine localizability conditions \cite{zhao2018affine,xu2020affine,zhu2022distributed,zhao2023specified,chen2020distributed,xiao2022framework,gao2022practical,li2020layered,wang2021affine,lin2021unified}, we firstly extend the affine localizability to non-generic nominal configuration. Although \cite{zhao2018affine} has pointed out this problem, the concrete affine localizability condition without generic assumption is not given.
	\item $(d + 1)$-rooted condition for directed graph is not necessary under EUC. It is proved for the first time that $(d + 1)$ in-neighbors condition can be weakened and extended to less in-neighbors while still maintaining the overall affine localizability.
	\item From perspective of construction method, we firstly establish a step-by-step construction method, i.e., EUC method, for achieving the nominal framework's affine localizability. Rather than only presenting overall conditions for nominal configuration and underlying interaction topology \cite{zhao2018affine,xu2020affine}, the internal connection mechanism of affine localizability is firstly revealed in this work.
	\item On basis of EUC method, the dynamical actions of adding and removing nodes  is firstly realized during affine formation maneuvering via flowable strategies proposed in Section III. This flexibility of formation size enables regulation of framework according to task requirements and environment constraints.
\end{enumerate}

\section{Fluidity of the Nominal Framework}
In this section, the preservation of affine localizability in cases of adding a new node and removing an old node are considered.
\begin{figure}[!t]
	\centering
	\includegraphics[width=3.2in]{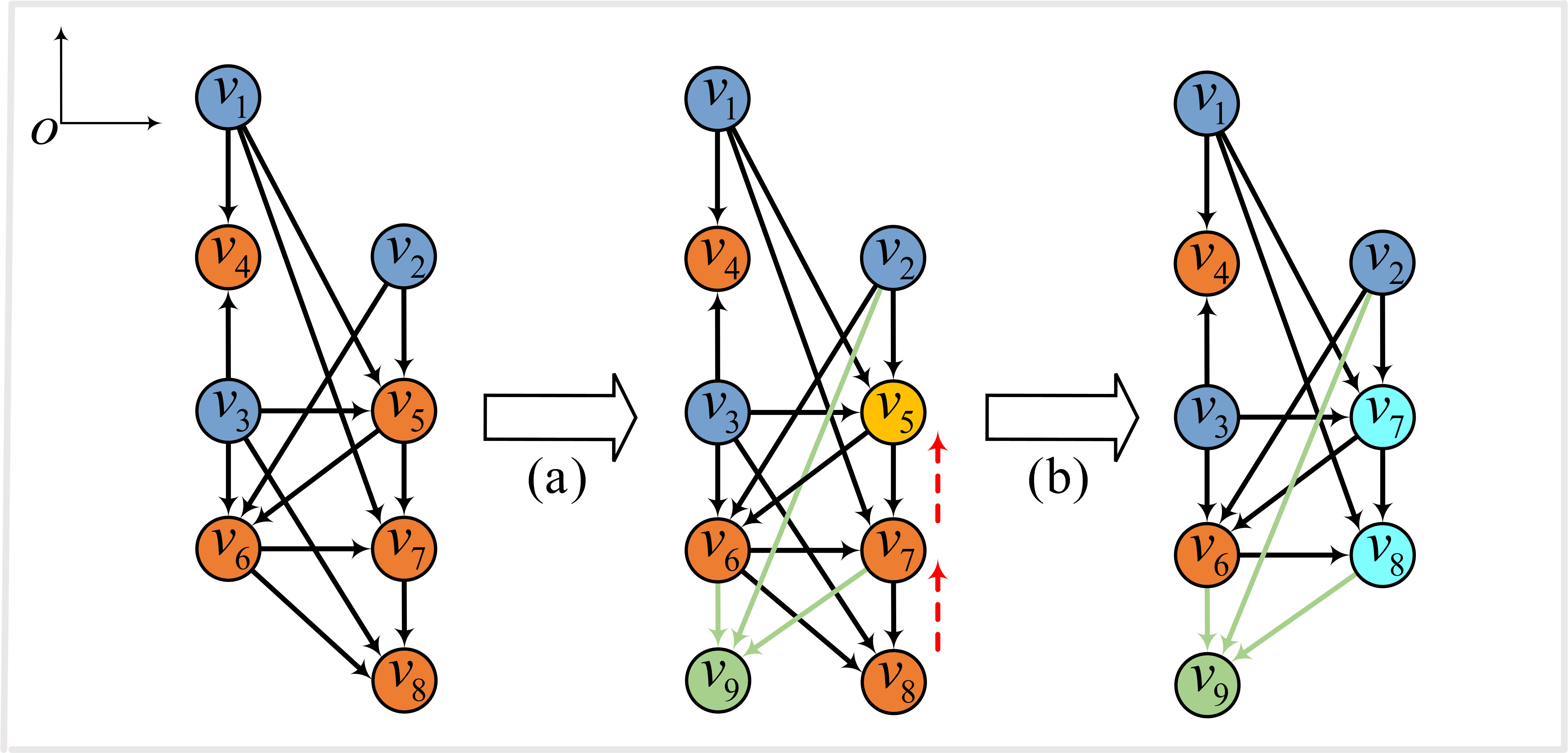}
	\caption{{(a) FIA of a new node $\mathcal{V}_{9}$ (green). (b) FOA of an old node $\mathcal{V}_{5}$ (yellow). Specifically, node $\mathcal{V}_{7}$ is shifted forward to replace node $\mathcal{V}_{5}$ and node $\mathcal{V}_{8}$ is shifted forward to replace $\mathcal{V}_{7}$.}}
	\label{flow}
\end{figure}

\subsection{Flow-in Action of New Node}
In this subsection, we address problem of how to add a new node to the existing nominal framework while guaranteeing the affine localizability of the reconstructed nominal framework.

Denoting the newly added node as $\mathcal{V}_{k}$, then we present the following result:

\noindent \textbf{Theorem 3.} \textit{(Affine localizability under flow-in action)} 
Given a nominal framework $(\mathcal{G},\chi)$ generated by EUC and a newly added node $\mathcal{V}_{k}$, we can maintain the affine localizability of the new overall nominal framework by the following flow-in action (FIA) steps:

\noindent \textit{Step 1.} Merge one $M_{d}$-equilibrium unit $R = \{r_{i}\}_{i = 0}^{M_{d}}$ with $2 \le M_{d} \le d + 1$ from $\mathcal{U}^{M_{d}}(\mathbb{R}^{d})$ into $(\mathcal{G},\chi)$ such that $\chi_{k}$ coincides with $r_{0}$ and $\{r_{i}\}_{i = 1}^{M_{d}}$ coincide with the existing points $\{\chi_{v_{i}}\}_{i = 1}^{M_{d}}$. Calculate the corresponding non-zero weights $\{w_{k,v_{j}}\}_{j=1}^{{M_{d}}}$ by \eqref{1} and \eqref{2}.

\noindent \textit{Step 2.} Reconstruct the nominal framework $(\mathcal{G}^{+},\chi^{+})$ with $\mathcal{G}^{+}=({\mathcal{V}}^{+},\mathcal{E}^{+}, \mathcal{W}^{+})$ by the following actions:
\begin{enumerate}
	\item Add new node $\mathcal{V}_{k}$ into the vertex set $\mathcal{V}^{+} = \mathcal{V} \cup \{\mathcal{V}_{k}\}$.
	\item Add new edges $(\mathcal{V}_{v_{j}},\mathcal{V}_{k})$, $j = 1, 2, \ldots, M_{d}$ into the edge set $\mathcal{E}^{+} = \mathcal{E} \cup \{(\mathcal{V}_{v_{j}},\mathcal{V}_{k})\}_{j=1}^{{M_{d}}}$.
	\item Add new non-zero weights $w_{k,v_{j}}$, $j = 1, 2, \ldots, M_{d}$ into the weight set $\mathcal{W}^{+} = \mathcal{W} \cup \{w_{k,v_{j}}\}_{j=1}^{{M_{d}}}$.
	\item Add $\chi_{k}$ into the nominal framework $\chi^{+} = \text{col}\{\chi,\chi_{k}\}$.
\end{enumerate}

The proof of \textit{Theorem 2} is omitted, here, because the FIA of a new node is the same operation as Step 2 in EUC. Thus, \textit{Theorem 1} can be used to verify the preservation of affine localizability.

\subsection{Flow-out Action of Old Node}
In this subsection, we address problem of how to remove an old node from the existing nominal framework while guaranteeing the affine localizability of the reconstructed nominal framework.

Denoting the removed node as $\mathcal{V}_{k}$, the affine localizability of the nominal framework is destroyed and needs to be repaired. To repair the and maintain the affine localizability, the following theorem can be derived:

\noindent \textbf{Theorem 4.} \textit{(Affine localizability under flow-out action)} 
Given a nominal framework $(\mathcal{G},\chi)$ generated by EUC. If an old node $\mathcal{V}_{k}$ is removed from the framework, we can maintain the affine localizability of the remaining nominal framework by the following flow-out action (FOA) steps:

\noindent \textit{Step 1.} If $\mathcal{V}_{k}$ is an end node, remove it and its associated edges, weights and position directly. If $\mathcal{V}_{k}$ is not an end node, let $\mathcal{V}_{k}$'s associated edges, weights and position in the framework be inherited by one of its out-neighbor $\mathcal{V}_{j}$. If $\mathcal{V}_{j}$ is not an end node, we can continue performing the inheritance operation until an end node is reached.

\noindent \textit{Step 2.} If $\mathcal{V}_{k}$ is an end node, reconstruct the nominal framework $(\mathcal{G}^{-},\chi^{-})$ with $\mathcal{G}^{-}=({\mathcal{V}}^{-},\mathcal{E}^{-}, \mathcal{W}^{-})$ by the following actions:
\begin{enumerate}
	\item Remove the old node $\mathcal{V}_{k}$ from the vertex set $\mathcal{V}^{-} = \mathcal{V} / \{\mathcal{V}_{k}\}$.
	\item Remove the old edges $(\mathcal{V}_{j},\mathcal{V}_{k})$, $j \in \mathcal{N}_{k}^{\text{in}}$ from the edge set $\mathcal{E}^{-} = \mathcal{E} / \{(\mathcal{V}_{j},\mathcal{V}_{k})\}_{j \in \mathcal{N}_{k}^{\text{in}}}$.
	\item Remove the old non-zero weights $w_{k,j}$, $j \in \mathcal{N}_{k}^{\text{in}}$ from the weight set $\mathcal{W}^{-} = \mathcal{W} / \{w_{k,j}\}_{j \in \mathcal{N}_{k}^{\text{in}}}$.
	\item Remove $\chi_{k}$ from the nominal framework $\chi^{-} = \text{col}\{\chi_{j}\}_{j \in \mathcal{V}^{-}}$.
\end{enumerate}
If $\mathcal{V}_{k}$ is not an end node, the inheritance operation is essentially equivalents to removing an end node, thus we can reconstruct the nominal framework by combining the above actions and inheritance operation.

\noindent \textit{Proof.} Please refer to \textit{Appendix E}.

\textit{Remark 4.} Examples of FIA of a new node and FOA of an old node are depicted in Fig. \ref{flow}. With the aid of our proposed FIA and FOA, we can add or remove  any node in or from the existing nominal framework without destroying the affine localizability of the reconstructed nominal framework. The prominent feature of the proposed flowable method lies in achieving the reconstruction procedure only via localized interaction, which will be discussed in Section IV.

\section{Localized Affine Formation Maneuver Control Law}
\begin{figure}[!h]
	\centering
	\includegraphics[width=3.2in]{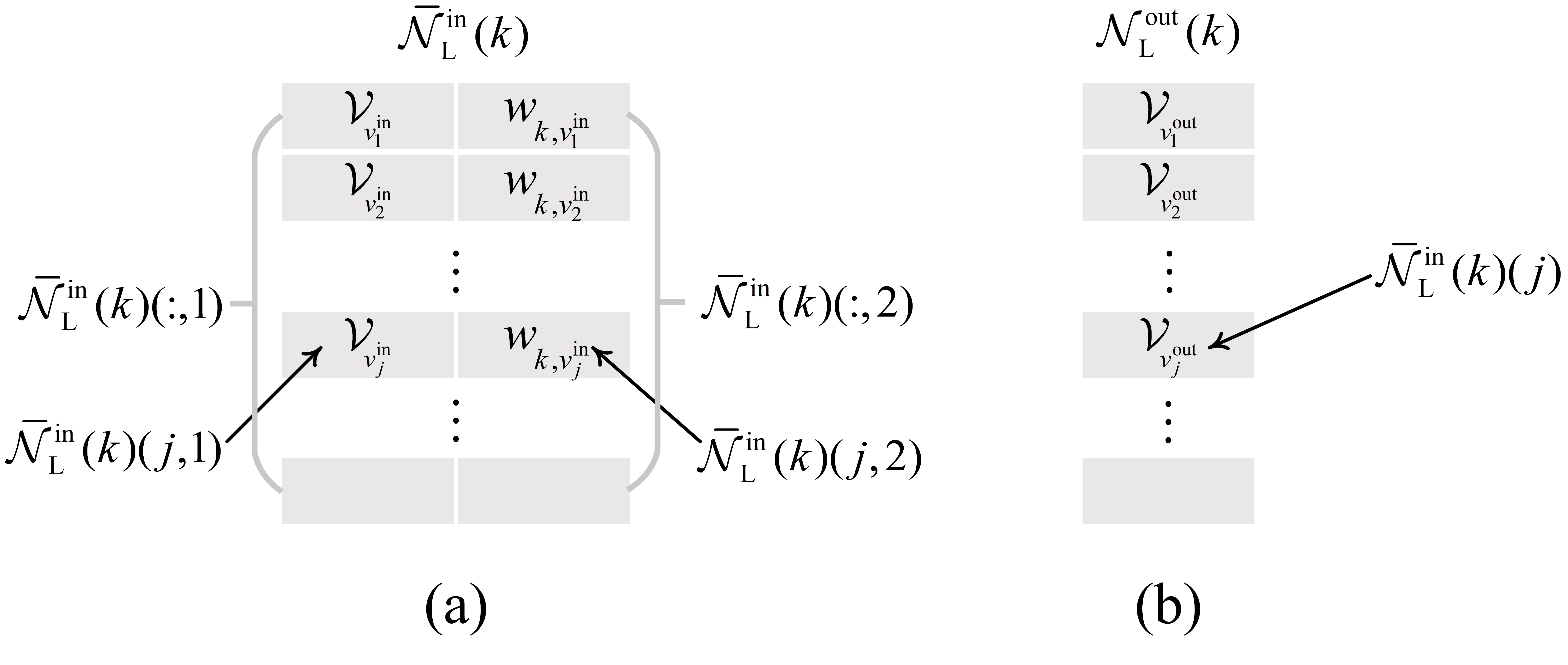}
	\caption{{(a) Data structure of ${\bar{\mathcal{N}}_{\text{L}}^{\text{in}}}(k)$. (b) Data structure of ${\mathcal{N}_{\text{L}}^{\text{out}}}(k)$.}}
	\label{data}
\end{figure}

In this section, we consider the localized updating of each node's stored information when FIA or FOA is performed. Besides, the design of localized affine formation maneuver control law is given subsequently. Before proceeding, we present the concept of local information of each follower node as follows:

\noindent \textbf{Definition 4.} \textit{(Local information)} Define the local information stored by node $\mathcal{V}_{k}$ as:
\begin{enumerate}
	\item ${\bar{\mathcal{N}}_{\text{L}}^{\text{in}}}(k)$: The collection of its in-neighbors and the associated non-zero weights.
	\item ${\mathcal{N}_{\text{L}}^{\text{out}}}(k)$: The collection of its out-neighbors.
\end{enumerate}
The data structures of local information are depicted in Fig. \ref{data}.

\noindent \textbf{Assumption 1.} The local information of each follower node is initialized by EUC described in \textit{Theorem 2}.

\subsection{Localized Communication Criterion}
\begin{algorithm} [!t]
	\caption{Operation of the added node $\mathcal{V}_{k}$}\label{Action of added node}
	\begin{algorithmic}[1]
		
		\State Initialize $\mathcal{V}_{k}$'s local information: $\bar{\mathcal{N}}_{\text{L}}^{\text{in}}(k)$, $\mathcal{N}_{\text{L}}^{\text{out}}(k)$;
		\State Backward propagation: Broadcast $\aleph^{1,\text{back}}$ to $\bar{\mathcal{N}}_{\text{L}}^{\text{in}}(k)(:,1)$;
		
	\end{algorithmic}
\end{algorithm}

\begin{algorithm} [!t]
	\caption{Action of node $\mathcal{V}_{j}$ which receives the backward propagation packet}\label{Action of backward}
	\begin{algorithmic}[1]
		
		\If{$\Psi_{\text{rec}} = -1$}
		\State Update $\mathcal{V}_{j}$'s local information: $\bar{\mathcal{N}}_{\text{L}}^{\text{in}}(j) \leftarrow \bar{\mathcal{N}}_{\text{L}}^{\text{in}}(j)$, $\mathcal{N}_{\text{L}}^{\text{out}}(j) \leftarrow \mathcal{N}_{\text{L}}^{\text{out}}(j)/\{\mathcal{V}_{\text{rec}}^{\text{self}}\}$;
		\Else
		\State Update $\mathcal{V}_{j}$'s local information: $\bar{\mathcal{N}}_{\text{L}}^{\text{in}}(j) \leftarrow \bar{\mathcal{N}}_{\text{L}}^{\text{in}}(j)$, $\mathcal{N}_{\text{L}}^{\text{out}}(j) \leftarrow \mathcal{N}_{\text{L}}^{\text{out}}(j)\cup \{\mathcal{V}_{\text{rec}}^{\text{self}}\}$;
		\EndIf
		
	\end{algorithmic}
\end{algorithm}

In this section, we consider the updating of the local information such that FIA and FOA can be performed successfully via localized communication. 

To construct the localized communication criterion (LCC), we first define the following two different types of information packet.

\noindent \textbf{Definition 5.} \textit{(Information packet)} An information packet is said to be a forward packet if it is broadcast from a node to its out-neighbors. On the contrary, an information packet is said to be a backward packet if it is broadcast from a node to its in-neighbors.

Now, we are ready to present two-type localized communication operations.

\noindent \textit{Step 1. (Forward propagation):} Forward propagation plays a role in informing one node's local information to its out-neighbors. A forward propagation packet is consisted of the following elements:
\begin{equation} \label{forward propagation}
	\aleph^{\text{for}}=\{\Psi = 1, \bar{{\mathcal{N}}}^{\text{in}}, {\mathcal{N}}^{\text{out}}, \mathcal{V}^{\text{self}}, \mathcal{H}^{\text{next}}, \mathcal{H}^{\text{pre}}\},
\end{equation}
where $\Psi$ is the direction identifier and $\Psi = 1$ means the forward identifier. $\bar{{\mathcal{N}}}^{\text{in}}$ and ${\mathcal{N}}^{\text{out}}$ are information identifiers. $\mathcal{V}^{\text{self}}$ is self-node identifier, $\mathcal{H}^{\text{next}}$ is the next inherit node identifier, and $\mathcal{H}^{\text{pre}}$ is the previous inherit node identifier.

\noindent \textit{Step 2. (Backward propagation):} Backward propagation plays a role in informing one node's in-neighbors about its requests. Two different backward propagation packet are consisted of the following elements:
\begin{equation} \label{backward propagation}
	\aleph^{1,\text{back}}=\{\Psi = -1, \mathcal{V}^{\text{self}}\},\aleph^{2,\text{back}}=\{\Psi = -2, \mathcal{V}^{\text{self}}\}.
\end{equation}
$\Psi = -1$ means the break backward identifier and $\Psi = -2$ means the link backward identifier.

Within this context, we can introduce the following localized updating algorithms. 

\noindent \textbf{LCC for FIA:} Assume that $\mathcal{V}_{k}$ is added. Then, for different nodes, we implement the following different operations.
\begin{enumerate}
	\item \textit{The added node $\mathcal{V}_{k}$:} Select $\mathcal{V}_{k}$'s in-neighbors. Calculate the corresponding none-zero weights associated with in-neighbors. Construct $\mathcal{V}_{k}$'s backward propagation packet as follows:
	\begin{align}
		\aleph^{2,\text{back}} \leftarrow \{-2,\mathcal{V}_{j}\}. \label{packet of added node}
	\end{align}
	Then, implement \textit{Algorithm \ref{Action of added node}} to broadcast and update information.
	\item \textit{For node $\mathcal{V}_{j}$ which receives the backward propagation packet:} Let $\aleph_{\text{rec}} \leftarrow \aleph^{2,\text{back}}$. then, implement \textit{Algorithm \ref{Action of backward}} to update information.
\end{enumerate}

\noindent \textbf{LCC for FOA:} Assume that $\mathcal{V}_{k}$ is removed. Then, for different nodes, we implement the following different operations.
\begin{enumerate}
	\item \textit{The removed node $\mathcal{V}_{k}$:} Select $\mathcal{V}_{k}$'s inherit node $\mathcal{H}_{k}^{\text{next}} \in \mathcal{N}_{\text{L}}^{\text{out}}(k)$. Construct $\mathcal{V}_{k}$'s forward propagation packet and backward propagation packet as follows:
	\begin{align}
		&\aleph^{\text{for}} \leftarrow \{1, \bar{\mathcal{N}}_{\text{L}}^{\text{in}}(k), \mathcal{N}_{\text{L}}^{\text{out}}(k), \mathcal{V}_{k}, \mathcal{H}_{k}^{\text{next}}, \varnothing\},\nonumber \\
		&\aleph^{1,\text{back}} \leftarrow \{-1,\mathcal{V}_{k}\}. \label{packet of removed node}
	\end{align}
	Then, implement \textit{Algorithm \ref{Action of removed node}} to broadcast and initialize information.
	\item \textit{For node $\mathcal{V}_{j}$ which receives the forward propagation packet:} Let $\aleph_{\text{rec}} \leftarrow \aleph^{\text{for}}$. If $\mathcal{V}_{j}$ is an inherit node, i.e., $\mathcal{V}_{j} =\mathcal{H}_{\text{rec}}^{\text{next}}$, then, select $\mathcal{V}_{j}$'s inherit node $\mathcal{H}_{j}^{\text{next}} \in \mathcal{N}_{\text{L}}^{\text{out}}(k)$ and construct $\mathcal{V}_{j}$'s forward propagation packet and backward propagation packet as follows:
	\begin{align}
		&\aleph^{\text{for}} \leftarrow \{1, \bar{\mathcal{N}}_{\text{L}}^{\text{in}}(j), \mathcal{N}_{\text{L}}^{\text{out}}(j), \mathcal{V}_{j}, \mathcal{H}_{j}^{\text{next}}, \mathcal{V}_{\text{rec}}^{\text{self}}\},\nonumber \\
		&\aleph^{1,\text{back}} \leftarrow \{-1,\mathcal{V}_{j}\}, \nonumber \\
		&\aleph^{2,\text{back}} \leftarrow \{-2,\mathcal{V}_{j}\}. \label{packet of inherit}
	\end{align}
	Then, implement \textit{Algorithm \ref{Action of inherit node}} to broadcast and update information. If $\mathcal{V}_{j}$ is not an inherit node, i.e., $\mathcal{V}_{j} \neq \mathcal{H}_{\text{rec}}^{\text{next}}$, then, implement \textit{Algorithm \ref{Action of non-inherit node}} to update information.
	\item \textit{For node $\mathcal{V}_{j}$ which receives the backward propagation packet:} Let $\aleph_{\text{rec}} \leftarrow \aleph^{1,\text{back}}$ ($\aleph_{\text{rec}} \leftarrow \aleph^{2,\text{back}}$). then, implement \textit{Algorithm \ref{Action of backward}} to update information.
\end{enumerate}

\textit{Remark 5.} With the help of our proposed algorithms, we can realize the FIA and FOA under LCC. Advantages of the proposed algorithms are summarized as follows:
\begin{enumerate}
	\item The proposed LCC has no reliance on decisions and solutions from the global perspective to achieve the FIA and FOA, i.e., only localized communication is needed to perform the reconstruction procedure.
	\item Compared with the existing topological reconfiguration methods \cite{zhang2023self}, we firstly establish feasible updating algorithms for each nodes via localized communication. On basis of our proposed LCC, the self-reconstruction of MASs can be realized by implementing the localized affine formation maneuver control law proposed in Section IV.C.
\end{enumerate}

\begin{algorithm} [!t]
	\caption{Operation of the removed node $\mathcal{V}_{k}$}\label{Action of removed node}
	\begin{algorithmic}[1]
		
		\State Forward propagation: Broadcast $\aleph_{\text{for}}$ to $\mathcal{N}_{\text{L}}^{\text{out}}(k)$;
		\State Backward propagation: Broadcast $\aleph^{1,\text{back}}$ to $\bar{\mathcal{N}}_{\text{L}}^{\text{in}}(k)$;
		\State Update $\mathcal{V}_{k}$'s local information: $\bar{\mathcal{N}}_{\text{L}}^{\text{in}}(k) \leftarrow \varnothing$, $\mathcal{N}_{\text{L}}^{\text{out}}(k) \leftarrow \varnothing$;
		
	\end{algorithmic}
\end{algorithm}

\begin{algorithm} [!t]
	\caption{Operation of the inherit node $\mathcal{V}_{j}$ which receives the forward propagation packet}\label{Action of inherit node}
	\begin{algorithmic}[1]
		
		\State Forward propagation: Broadcast $\aleph^{\text{for}}$ to $\mathcal{N}_{\text{L}}^{\text{out}}(j)$;
		\State Backward propagation: Broadcast $\aleph^{1,\text{back}}$ to $\mathcal{N}_{\text{L}}^{\text{in}}(j)(:,1)$;
		\If{$\mathcal{H}_{\text{rec}}^{\text{pre}} = \varnothing$}
		\State Update $\mathcal{V}_{j}$'s local information: $\bar{\mathcal{N}}_{\text{L}}^{\text{in}}(j) \leftarrow \bar{\mathcal{N}}_{\text{rec}}^{\text{in}}$, $\mathcal{N}_{\text{L}}^{\text{out}}(j) \leftarrow ({\mathcal{N}}_{\text{rec}}^{\text{out}}:\mathcal{V}_{j} \leftarrow \mathcal{H}_{j}^{\text{next}})$;
		\State Backward propagation: Broadcast $\aleph^{2,\text{back}}$ to $\mathcal{N}_{\text{L}}^{\text{in}}(j)(:,1)$;
		\Else
		\State Update $\mathcal{V}_{j}$'s local information: $\bar{\mathcal{N}}_{\text{L}}^{\text{in}}(j)(:,1) \leftarrow (\bar{\mathcal{N}}_{\text{rec}}^{\text{in}}(:,1):\mathcal{H}_{\text{rec}}^{\text{pre}} \leftarrow \mathcal{V}_{\text{rec}}^{\text{self}})$, $\bar{\mathcal{N}}_{\text{L}}^{\text{in}}(j)(:,2) \leftarrow \bar{\mathcal{N}}_{\text{rec}}^{\text{in}}(:,2)$,  $\mathcal{N}_{j}^{\text{out}}(\text{L}) \leftarrow 	({\mathcal{N}}_{\text{rec}}^{\text{out}}:\mathcal{V}_{j} \leftarrow \mathcal{H}_{j}^{\text{next}})$;
		\State Backward propagation: Broadcast $\aleph^{2,\text{back}}$ to $\bar{\mathcal{N}}_{\text{L}}^{\text{in}}(j)/\{\mathcal{V}_{\text{rec}}^{\text{self}}\}$;
		\EndIf
		
	\end{algorithmic}
\end{algorithm}

\begin{algorithm} [!t]
	\caption{Operation of non-inherit node $\mathcal{V}_{j}$ which receives the forward propagation packet}\label{Action of non-inherit node}
	\begin{algorithmic}[1]
		
		\State Update $\mathcal{V}_{j}$'s local information: $\bar{\mathcal{N}}_{\text{L}}^{\text{in}}(j)(:,1) \leftarrow (\bar{\mathcal{N}}_{\text{L}}^{\text{in}}(j)(:,1):\mathcal{V}_{\text{rec}}^{\text{self}} \leftarrow \mathcal{H}_{\text{rec}}^{\text{next}})$, $\bar{\mathcal{N}}_{\text{L}}^{\text{in}}(j)(:,2) \leftarrow \bar{\mathcal{N}}_{\text{L}}^{\text{in}}(j)(:,2)$, $\mathcal{N}_{\text{L}}^{\text{out}}(j) \leftarrow \mathcal{N}_{\text{L}}^{\text{out}}(j)$;
		
	\end{algorithmic}
\end{algorithm}

\subsection{Existence of Stabilizing Control Gains}
Time sequence of adding and removing nodes to the existing nominal framework is defined as $\{T_1, T_2, \ldots, T_{k}, \ldots\}$, where $T_k$ denotes the time instant that a new node (an old node) is added to (removed from) the existing nominal framework. Then, let $\Gamma^{T} = \{T_{0}=0, T_{1}, \ldots, T_{k}, \ldots \}$, we define $\mathcal{V}(k) = \mathcal{V}_{l} \cup \mathcal{V}_{f}(k)$ as the vertex set corresponding to time interval $[T_{k},T_{k + 1})$, $k \in \mathbb{N}$. The sequences of edge set and weight set are defined as $\mathcal{E}(k)$ and $\mathcal{W}(k)$. For $\mathcal{V}_{i} \in \mathcal{V}(k)$, $\mathcal{N}_{i}^{\text{in}}(k)$ and $\mathcal{N}_{i}^{\text{out}}(k)$ denote the in-neighbor set and out-neighbor set of $\mathcal{V}_{i}$ during time interval $[T_{k},T_{k + 1})$. For $\mathcal{V}_{i} \in \mathcal{V}(k)$, $\chi_{i}(k)$ denotes the nominal position of $\mathcal{V}_{i}$ during time interval $[T_{k},T_{k + 1})$. Thus, the sequence of nominal framework is formed as $(\mathcal{G}(k), \chi(k))$ with $\mathcal{G}(k) = (\mathcal{V}(k), \mathcal{E}(k), \mathcal{W}(k))$.

\noindent \textbf{Assumption 2.} The sequences of nominal framework is initialized by EUC described in \textit{Theorem 1}, i.e., $\mathcal{V}(0) = \mathcal{V}$, $\mathcal{E}(0) = \mathcal{E}$, $\mathcal{W}(0) = \mathcal{W}$ and $\chi(0) = \chi$.

This section brief studies the $n$-th order integrator dynamics. Thus, agent $\mathcal{V}_{i}$ are governed as
\begin{equation} \label{model 1}
	\left\{ \begin{aligned}
		\dot{x}_{k,i} &= x_{k+1,i}, \text{ } x_{1,i} = p_{i} \in\mathbb{R}^{d} \\
		\dot{x}_{n,i} &= u_{i}, \text{ } k=1,2,\ldots,n-1
	\end{aligned}
	\right.,
\end{equation}
where $x_{i}\in\mathbb{R}^{nd}$ and $u_{i}\in\mathbb{R}^{d}$ are the state and control input, respectively, $p_{i}$ denotes the position. 

\noindent \textbf{Assumption 3.} Leader trajectories are $n$-th order differentiable and set as $p_{i}(t) = p^{*}_{i}(t) = A(t)\chi_{i}+ b(t)= \sum_{k=0}^{n-1}c_{k,i}t^{k}, \forall i\in\mathcal{V}_{l}$ with $c_{k,i}\in\mathbb{R}^{d}$.

Next, we aim to confirm the existence of stabilizing control gains. To facilitate the following analysis, define $C\in\mathbb{R}^{n\times n}$ and $D\in\mathbb{R}^{n}$ as
\begin{equation} \label{state matrix and input matrix}
	C = \begin{bmatrix}
		0_{n-1} & I_{n-1} \\
		0 & 0^{\text{T}}_{n-1}
	\end{bmatrix}, \text{ } D = \begin{bmatrix}
		0_{n-1} \\
		1
	\end{bmatrix}.
\end{equation}
Meanwhile,  for brevity, we define
\begin{equation}\label{breif}
\begin{aligned}
& \bar{C}=I_{|\mathcal{V}_{f}(k)|}\otimes C \otimes I_{d}, \text{ } \bar{D} = I_{|\mathcal{V}_{f}(k)|} \otimes D \otimes I_{d}. \\
\end{aligned} 
\end{equation}

\noindent \textbf{Lemma 4.} \cite{zhao2023specified} For invertible matrix $\Omega_{ff}(k)$, $k \in \mathbb{N}$, there exist diagonal matrix $Q(k)$ such that $Q(k)\Omega_{ff}(k) + \Omega_{ff}^{\text{T}}(k)Q(k)\succ 0$.

On the other hand, owing to $\text{Rank}([D, CD]) = n$, $\left(C, D\right)$ is controllable. Thus, there exist  positive-definite solutions $P\in\mathbb{R}^{n\times n}$ to the following Riccati equation:
\begin{equation} \label{PLE}
	C^{\text{T}}P + PC - PDD^{\text{T}}P  = -\xi P,
\end{equation}
where $\xi$ is a positive constant.

\noindent \textbf{Lemma 5.} \cite{zhou2024affine} {Let $T^{*} > 0$ be a designed constant. For given constants $\varepsilon \in (0,1)$ and $\mu > 0$. Select $\bar \Gamma(T,\varepsilon,\mu)$ as}
{\begin{equation}
		\label{G5_1}
		\begin{aligned}
			\bar \Gamma(T^{*},\varepsilon,\mu) = \frac{1}{T^{*} \sqrt{4\mu + \mu^2}} \ln (\frac{\Gamma_1(\varepsilon,\mu)\Gamma_4(\varepsilon,\mu) }{\Gamma_3(\varepsilon,\mu)\Gamma_2(\varepsilon,\mu) }),
		\end{aligned}
\end{equation}}{with}
{\begin{equation}
		\label{G5_2}
		\begin{aligned}
			&\Gamma_1(\varepsilon,\mu)=\frac{2+\mu}{{2}} + \frac{\sqrt {4\mu + \mu^2}}{2} + \varepsilon,\\
			&\Gamma_2(\varepsilon,\mu)=\frac{2+\mu}{{2}} - \frac{\sqrt {4\mu + \mu^2}}{2} + \varepsilon,\\	
			&\Gamma_3(\varepsilon,\mu)=1 + (\frac{2+\mu}{{2}} + \frac{\sqrt {4\mu + \mu^2}}{2})\varepsilon,\\
			&\Gamma_4(\varepsilon,\mu)=1 + (\frac{2+\mu}{{2}} - \frac{\sqrt {4\mu + \mu^2}}{2})\varepsilon.
		\end{aligned}
\end{equation}}{Consider the auxiliary variable $\underline{\phi}:\mathbb{R}_{\geq 0} \longrightarrow \mathbb{R}$ defined by}
{\begin{equation}
		\label{G6}
		\resizebox{0.885\hsize}{!}{$
			\begin{aligned}
				\underline{\phi} (t) = \frac{\sqrt {4\mu + \mu^2}}{2}  \frac{{\Gamma_3(\varepsilon,\mu)  + \Gamma_4(\varepsilon,\mu){e^{ - \bar \Gamma(T^{*},\varepsilon,\mu) \sqrt {4\mu + \mu^2} t}}}}{{\Gamma_3(\varepsilon,\mu) - \Gamma_4(\varepsilon,\mu){e^{ - \bar \Gamma(T^{*},\varepsilon,\mu) \sqrt {4\mu + \mu^2} t}}}} - \frac{2+\mu}{{2}}.
			\end{aligned}$}
\end{equation}}{Then, $\underline{\phi}(t)$ is monotonically decreasing and satisfies}
{\begin{align}
			\underline{\dot \phi}(t)  =  - \bar \Gamma(T^{*},\varepsilon,\mu)({\underline{\phi} ^2(t)} + (2+\mu)\underline{\phi}(t)  + 1),\label{G7}
\end{align}}{with $\underline{\phi} (0) = 1/{\varepsilon }$, $\underline{\phi} ({T^{*}}) = \varepsilon $.}

\subsection{Localized Sensing Based Affine Formation Maneuver Control}
Let $\Gamma^{\Delta} = \{\Delta_{0} = 0, \Delta_{1}, \ldots, \Delta_{m}, \ldots \}$ be the sequence of measurement switching instants (SMSI) with $ \Delta_{m}, m\in\mathbb{Z}$ the $m$-th switched instant.

The associated tracking error (TE) $e_{i}(t)\in\mathbb{R}^{nd}$, $t \in [\Delta_{m},\Delta_{m+1})$ is described as
\begin{equation} \label{model 2}
	e_{1,i} = x_{1,i} - \hat{p}_{i}^{*}, \text{ } e_{k,i} = x_{k,i} - \mathbb{D}^{k-1}\hat{p}^{*}_{i}, \text{ }k=2,\ldots, n,
\end{equation}
where $\hat{p}_{i}^{*}$ is defined as
\begin{equation} \label{model 2_2}
	\begin{aligned}
&\hat{p}_{i}^{*}(t) = A(t)\chi_{i}(k(m)) + b(t), \\
&T_{k({m})} = \text{sup}\{t \le \Delta_{m} | t \in \Gamma^{T}\}.
	\end{aligned}
\end{equation}

We proposed the following LSAFMC protocol for follower $\mathcal{V}_i$
\begin{align}
u_{i}(t) = - \beta q_i(k(m)) (\gamma^{\text{T}} \otimes I_d)\hat{s}_i(t), \text{ } t \in [\Delta_{m},\Delta_{m+1}), \label{control scheme 1}
\end{align}
with $\beta > 0$ and $\gamma = \text{col}\{\gamma_i\}_{i = 1,2,...,n}$ being positive scales to be set. Where $\hat{s}_i(t)$ is an available weighted summation of relative states (WSRE) derived from ${s}_i(t)$
\begin{equation}
\label{control scheme 2}
\begin{aligned}
&\begin{cases}
\hat{s}_{i}(t) = s_{i}(t), \text{ } t = \Delta_{m} \\
\dot{\hat{s}}_{i}(t) = (C \otimes I_{d}) \hat{s}_{i}(t), \text{ } t \in [\Delta_{m}, \Delta_{m+1})
\end{cases},
\end{aligned}
\end{equation}
with $s_{i}(t)\in\mathbb{R}^{dn}$ defined as
\begin{equation} \label{tracking error 1}
	\begin{aligned}
	&s_{i}(t) = \sum_{j\in\mathcal{N}^{\text{in}}_{i}(k({m}))}\omega_{i,j}(k({m})) \left(x_{i}(t)-x_{j}(t)\right).
	\end{aligned}
\end{equation}
WSRE has the potential to transform distributed control problems into a local coordinate system, simplifying the complexity of stability analysis. Then, the discrepancy between WSRE and the available WSRE is expressed as
\begin{align}
\delta_{i}\left(t\right) = {s}_{i}\left(t\right) - \hat{s}_{i}\left(t\right). \label{control scheme 3}
\end{align}
Based on the above discussion and , we are able to give the main results of this section.

\vspace{0.05in}
\noindent \textbf{Theorem 5.} Consider LSAFMC-based MASs \eqref{model 1} obeying \textit{Assumptions 1-3}. If the SMSI is designed as $\Delta_{m+1} - \Delta_{m} \le T^{*}, m\in\mathbb{N}$, $P$ is a positive-definite solution to \eqref{PLE} with $\xi$ is chosen such that
\begin{align} 
&\epsilon = \frac{\lambda_{\text{max}}(PDD^{\text{T}}P)}{\lambda_{\text{min}}(P)} \le \min_{k\in\mathbb{N}}\left\{\mathcal{Z}_{1}, \mathcal{Z}_{2}, \mathcal{Z}_{3}\right\}, \nonumber\\
&\mathcal{Z}_{1} = \bar \Gamma(T^{*},\varepsilon,\mu)\frac{1}{l_{1}\beta^2\lambda_{\text{max}}^{2}(Q(k)\Omega_{ff}(k))}, \nonumber\\
&\mathcal{Z}_{2} = \bar \Gamma(T^{*},\varepsilon,\mu)\frac{1}{l_{2}\beta^{2}}, \label{control gains 1}\\
&\mathcal{Z}_{3} = \frac{2\bar \Gamma(T^{*},\varepsilon,\mu)}{1 + \beta \lambda_{\text{max}}(Q(k)\Omega_{ff}(k)+\Omega_{ff}^{\text{T}}(k)Q(k))}, \nonumber
\end{align}
and control gains are set as 
\begin{align}
&\qquad \qquad \qquad \qquad \quad \gamma^{\text{T}} = D^{\text{T}} P, \label{control gains 2} \\
&\beta \ge \frac{1 + \frac{1}{l_1} + \frac{\lambda_{\text{max}}^{2}(Q(k)\Omega_{ff}(k))}{l_2}}{\lambda_{\text{min}}(Q(k)\Omega_{ff}(k)+\Omega_{ff}^{\text{T}}(k)Q(k))}, \text{ } k\in\mathbb{N}, \nonumber
\end{align}
where $l_1,l_2 > 0$. Then, there exist a positive constant $\eta$ satisfies $\eta < \xi$ and $\sigma > 0$ such that the following functions $\theta_i(t)$ and $g_i(t)$
\begin{align}
\theta_i(t) = \mathcal{T}_{m,i}e^{(\xi - \eta)(t - \Delta_{m})}, \text{ } t \in [\Delta_{m}, \Delta_{m+1}),\label{control gains 3}
\end{align}
\begin{align}
	g_i(t) = \mathcal{S}_{m,i}e^{-\sigma(t - \Delta_{m})}, \text{ } t \in [\Delta_{m}, \Delta_{m+1}),\label{control gains 4}
\end{align}
guarantees the exponentially stable of MASs. $\Gamma^{\mathcal{T}}_{i} = \{\mathcal{T}_{0,i}, \mathcal{T}_{1,i}, \ldots, \mathcal{T}_{m,i}, \ldots \}$ and $\Gamma^{\mathcal{S}}_{i} = \{\mathcal{S}_{0,i}, \mathcal{S}_{1,i}, \ldots, \mathcal{S}_{m,i}, \ldots \}$ are determined in the subsequent proof procedure.

\noindent \textit{Proof.} From \eqref{tracking error 1}, for $t \in [\Delta_{m}, \Delta_{m+1})$, we can derive
\begin{align}
s_{f}(t) = (\Omega_{fl}(k({m})) \otimes I_{nd})x_{l}(t) + (\Omega_{ff}(k({m})) \otimes I_{nd})s_{f}(t). \label{tracking error 2}
\end{align}
Under \textit{Assumption 2}, the underlying nominal framework $(\mathcal{G},\chi)$ is affinely localizable. Then, we can derive from \eqref{model 2} and \eqref{tracking error 2} that
\begin{equation} \label{tracking error new 2}
s_{f}(t) = (\Omega_{ff}(k({m})) \otimes I_{nd}) e_{f}(t).
\end{equation}
From \eqref{state matrix and input matrix}, $C$ is a backward shift matrix, thus $(\Omega_{fl}(k({m})) \otimes C \otimes I_{d} )x_{l}(t) + (\Omega_{ff}(k({m})) \otimes C \otimes I_{d} )s_{f}(t) = ( I_{|\mathcal{V}_{f}(k({m}))|} \otimes C \otimes I_{d} ) ((\Omega_{fl}(k({m})) \otimes I_{nd})x_{l}(t) + (\Omega_{ff}(k({m})) \otimes I_{nd}) s_{f}(t))$. Then, by means of \eqref{tracking error 2}, the tracking dynamics under \textit{Assumption 3} can be derived as
\begin{align}
\dot{s}_{f}(t) = \bar{C} {s}_{f}(t) + (\Omega_{ff}(k({m})) \otimes I_{nd}) \bar{D} {u}_{f}(t). \label{B1}
\end{align}
Since the measurement is not continuous, from \eqref{control scheme 1}-\eqref{control scheme 3}, the control input can be rewritten as $u_{i}(t) = -\beta ( \gamma^{\text{T}} \otimes I_{d} ) ( s_{i} - \delta_{i})$. The stack of control inputs is
\begin{align} 
{u}_{f}(t) = - \beta \bar Q(k({m})) \bar{\gamma}^{\text{T}} s_{f}(t) + \beta \bar Q(k({m})) \bar{\gamma}^{\text{T}} \delta_{f}(t), \label{B2}
\end{align}
where $\bar{\gamma} = \gamma \otimes I_{d}$ and $\bar Q(k({m})) = Q(k({m})) \otimes I_{nd}$. Then, we set the Lyapunov candidate as
\begin{align}
V(t, s_{f}(t), \delta_{f}(t)) =  V_1(t, s_{f}(t)) + V_2(t, \delta_{f}(t)) + V_{3}(t). \label{B3}
\end{align}
When $t \in [\Delta_{m}, \Delta_{m+1})$, $V_1$, $V_2$, and $V_3$ are defined as
\begin{align}
&V_1 = s^{\text{T}}_{f}(t)\bar \theta(t) \bar P s_{f}(t),\nonumber\\
&V_2 = \phi \delta^{\text{T}}_{f}(t)\bar \theta(t) \bar P \delta_{f}(t),\nonumber\\
&V_3 = \sum_{i \in \mathcal{V}_{f}(k({m}))}g_i(t), \label{B4} 
\end{align}
where $\bar \theta(t)  = \text{diag}\{\theta_f(t)\} \otimes I_{nd}$ and $\bar P = I_{|\mathcal{V}_{f}(k({m}))|} \otimes P \otimes I_{d}$. $\phi$ is derived from $\underline{\phi}$
\begin{align}
\phi(t) = \underline{\phi}(t-\Delta_{m}), \text{ } t \in [\Delta_{m}, \Delta_{m+1}).\label{B4_2} 
\end{align}

1) Firstly, considering the flow dynamics of $V$, i.e., $t \in [\Delta_{m}, \Delta_{m+1})$, we have
\begin{equation} \label{B5}
\begin{aligned}
&\dot{V}_1 = \dot{s}^{\text{T}}_{f}(t)\bar \theta(t) \bar P s_{f}(t) \\
&+ s^{\text{T}}_{f}(t)\bar \theta(t) \bar P \dot{s}_{f}(t) +s^{\text{T}}_{f}(t)\dot{\bar \theta}(t) \bar P s_{f}(t),
\end{aligned}
\end{equation}
Since $\bar \theta(t)$ and $\bar P$ are symmetric, we derive from \textit{Lemma 3} that $\text{diag}\{\theta_f(t)\}Q(k({m}))\Omega_{ff}(k({m}))+\Omega_{ff}^{\text{T}}(k({m}))Q(k({m}))\text{diag}\{\theta_f(t)\} \ge \lambda_{\text{min}}(Q(k({m}))\Omega_{ff}(k({m}))+\Omega_{ff}^{\text{T}}(k({m}))Q(k({m})))\text{diag}\{\theta_f(t)\}$. Together with \eqref{control gains 2}, we can derive
\begin{align}
&\dot{s}^{\text{T}}_{f}(t)\bar \theta(t) \bar P s_{f}(t) + s^{\text{T}}_{f}(t)\bar \theta(t) \bar P \dot{s}_{f}(t) \le \label{B6}\\
&2s^{\text{T}}_{f}(t)\bar \theta(t) \bar P \bar Cs_{f}(t)  - \beta \varpi_1 s^{\text{T}}_{f}(t)\bar \theta(t) \bar P \bar D \bar D^{\text{T}} \bar Ps_{f}(t) \nonumber\\
& + \varpi_2 s^{\text{T}}_{f}(t)\bar \theta(t) \bar P \bar D \bar D^{\text{T}} \bar Ps_{f}(t) + \varpi_3 \delta^{\text{T}}_{f}(t)\bar \theta(t) \bar P \delta_{f}(t),\nonumber
\end{align}
where $\varpi_1 = \lambda_{\text{min}}(Q(k({m}))\Omega_{ff}(k({m}))+\Omega_{ff}^{\text{T}}(k({m}))Q(k({m})))$, $\varpi_2 = 1 / l_1$, and $\varpi_3 = l_1\beta^{2} \lambda_{\text{max}}^{2}(Q(k({m}))\Omega_{ff}(k({m}))) \epsilon$. In light of \eqref{PLE} and \eqref{control gains 3}, we have 
\begin{align}
\dot{V}_1 \le& -\eta s^{\text{T}}_{f}(t)\bar \theta(t) \bar P s_{f}(t) \label{B7}\\
& - \varpi_4 s^{\text{T}}_{f}(t)\bar \theta(t) \bar P \bar D \bar D^{\text{T}} \bar Ps_{f}(t) \nonumber\\
& + \varpi_3 \delta^{\text{T}}_{f}(t)\bar \theta(t) \bar P \delta_{f}(t),  \nonumber
\end{align}
where $\varpi_4 = \beta \varpi_1 - \varpi_2 - 1$. Then, we conduct $\dot{V}_{2}$ as follows
\begin{equation} \label{B8}
	\begin{aligned}
		\dot{V}_2 =& \dot{\phi}\delta^{\text{T}}_{f}(t)\bar \theta(t) \bar P \delta_{f}(t) + \phi \dot{\delta}^{\text{T}}_{f}(t) \bar \theta(t) \bar P \delta_{f}(t) \\
		&+ \phi \delta^{\text{T}}_{f}(t)\bar \theta(t) \bar P \dot{\delta}_{f}(t) +\phi \delta^{\text{T}}_{f}(t)\dot{\bar \theta}(t) \bar P \delta_{f}(t).
	\end{aligned}
\end{equation}
By recalling \eqref{control scheme 2} and \eqref{B1}, we have
\begin{align}
	\dot{\delta}_{f}(t) = \bar{C} \delta_{f}(t) + (\Omega_{ff}(t) \otimes I_{nd}) \bar{D} u_{f}(t). \label{B9}
\end{align}
Similar to \eqref{B6} and \eqref{B7}, we recall \eqref{PLE} and \eqref{control gains 3} and use \textit{Young's inequality} to yield
\begin{align}
&\phi \dot{\delta}^{\text{T}}_{f}(t)\bar \theta(t) \bar P \delta_{f}(t) + \phi \delta^{\text{T}}_{f}(t)\bar \theta(t) \bar P \dot{\delta}_{f}(t)  \nonumber\\
&+\phi \delta^{\text{T}}_{f}(t)\dot{\bar \theta}(t) \bar P \delta_{f}(t)  \le  \label{B10}\\
&[\varpi_5 \epsilon \phi^{2} + (\epsilon + \varpi_6 \epsilon - \eta)\phi] \delta^{\text{T}}_{f}(t)\bar \theta(t) \bar P \delta_{f}(t) \nonumber\\
&+ \varpi_7 s^{\text{T}}_{f}(t)\bar \theta(t) \bar P \bar D \bar D^{\text{T}} \bar Ps_{f}(t),
\end{align}
where $\varpi_5 = l_2\beta^{2}$, $\varpi_6 = \beta \lambda_{\text{max}}(Q(k({m}))\Omega_{ff}(k({m}))+\Omega_{ff}^{\text{T}}(k({m}))Q(k({m})))$, and $\varpi_7 = \lambda_{\text{max}}^{2}(Q(k)\Omega_{ff}(k({m})))/l_2$. Then, by means of \eqref{control gains 1} and \eqref{control gains 2} and recalling \textit{Lemma 5}, we can derive
\begin{align}
\dot{V}_1 + \dot{V}_2 \le -\eta V_1 - \mu \bar \Gamma(T^{*},\varepsilon,\mu) V_2. \label{B11}
\end{align}
Finally, we differentiate $V_3$ as follows
\begin{align}
	\dot{V}_3 \le -\sigma V_3. \label{B12}
\end{align}
By combining \eqref{B11} and \eqref{B12}, we have
\begin{align}
	\dot{V} \le -\min\{\eta, \mu \bar \Gamma(T^{*},\varepsilon,\mu), \sigma\}V. \label{B13}
\end{align}

2) Secondly, considering the jump dynamics of $V$, i.e., $t = \Delta_{m+1}$.

\noindent Case 1: No follower node is added and removed during time interval $[\Delta_{m}, \Delta_{m+1})$. Then, $\mathcal{T}_{m + 1,i}$ and $\mathcal{S}_{m + 1,i}$, $\mathcal{V}_i \in \mathcal{V}_{f}(k({m}+1))$ are designed as follows
\begin{align}
\mathcal{T}_{m + 1,i} = \theta_i(\Delta_{m + 1}^{-}), \text{ } \mathcal{S}_{m + 1,i} = g_i(\Delta_{m + 1}^{-}). \label{B14}
\end{align}
Thus, from \eqref{B14}, we have
\begin{align}
	V(\Delta_{m + 1}) - V(\Delta^{-}_{m + 1}) = 0. \label{B15}
\end{align}

\noindent Case 2: One node is added to be a new follower during time interval $[\Delta_{m}, \Delta_{m+1})$. Assume that $\mathcal{V}_{k}$ is the newly flow-in node, then $\mathcal{T}_{m + 1,i}$ and $\mathcal{S}_{m + 1,i}$, $\mathcal{V}_i \in \mathcal{V}_{f}(k({m}+1))$ are designed as follows
\begin{align}
	&\mathcal{T}_{m + 1,k}= \nonumber\\
	& \frac{1}{q_k(k({m}+1))s_k^{\text{T}}(\Delta_{m + 1})(P \otimes I_{d})s_k(\Delta_{m + 1}) + \hbar_1}\nonumber\\
	&\times \frac{1}{2\hbar_2}\sum_{\mathcal{V}_i \in \mathcal{V}_{f}(k({m}))}g_i(\Delta_{m + 1}^{-}),\nonumber\\
	&\mathcal{S}_{m + 1,k} = \frac{1}{2\hbar_2}\sum_{\mathcal{V}_i \in \mathcal{V}_{f}(k({m}))}g_i(\Delta_{m + 1}^{-}),\nonumber\\
	&\mathcal{T}_{m + 1,i} = \theta_i(\Delta_{m + 1}^{-}),\text{ }\mathcal{V}_i \in \mathcal{V}_{f}(k({m})),\nonumber\\
	&\mathcal{S}_{m + 1,i} = g_i(\Delta_{m + 1}^{-}) - \frac{1}{\hbar_2}g_i(\Delta_{m + 1}^{-}), \text{ }\mathcal{V}_i \in \mathcal{V}_{f}(k({m})). \label{B18}
\end{align}
Thus, from \eqref{B16}, we have
\begin{align}
	V(\Delta_{m + 1}) - V(\Delta^{-}_{m + 1}) \le 0. \label{B18}
\end{align}

\noindent Case 3: One of the follower nodes is removed during time interval $[\Delta_{m}, \Delta_{m+1})$. Then $\mathcal{T}_{m + 1,i}$ and $\mathcal{S}_{m + 1,i}$, $\mathcal{V}_i \in \mathcal{V}_{f}(k({m}+1))$ are designed as follows
\begin{align}
	&\mathcal{T}_{m + 1,i}= \nonumber\\
	& \frac{1}{q_i(k({m}+1))s_i^{\text{T}}(\Delta_{m + 1})(P \otimes I_{d})s_i(\Delta_{m + 1}) + \hbar_1}\frac{1}{\hbar_2}g_i(\Delta_{m + 1}^{-}),\nonumber\\
	&\mathcal{S}_{m + 1,i} = g_i(\Delta_{m + 1}^{-}) - \frac{1}{\hbar_2}g_i(\Delta_{m + 1}^{-}).\label{B16}
\end{align}
Thus, from \eqref{B16}, we have
\begin{align}
	V(\Delta_{m + 1}) - V(\Delta^{-}_{m+1}) \le 0. \label{B17}
\end{align}

By combining \eqref{B13}, \eqref{B15}, \eqref{B18}, \eqref{B17}, and \eqref{tracking error new 2}, we can then ensure that $e_{f}(t)$ is exponentially stable. This completes the proof. $\blacksquare$

\begin{figure}[!t]
	\centering
	\includegraphics[width=3.2in]{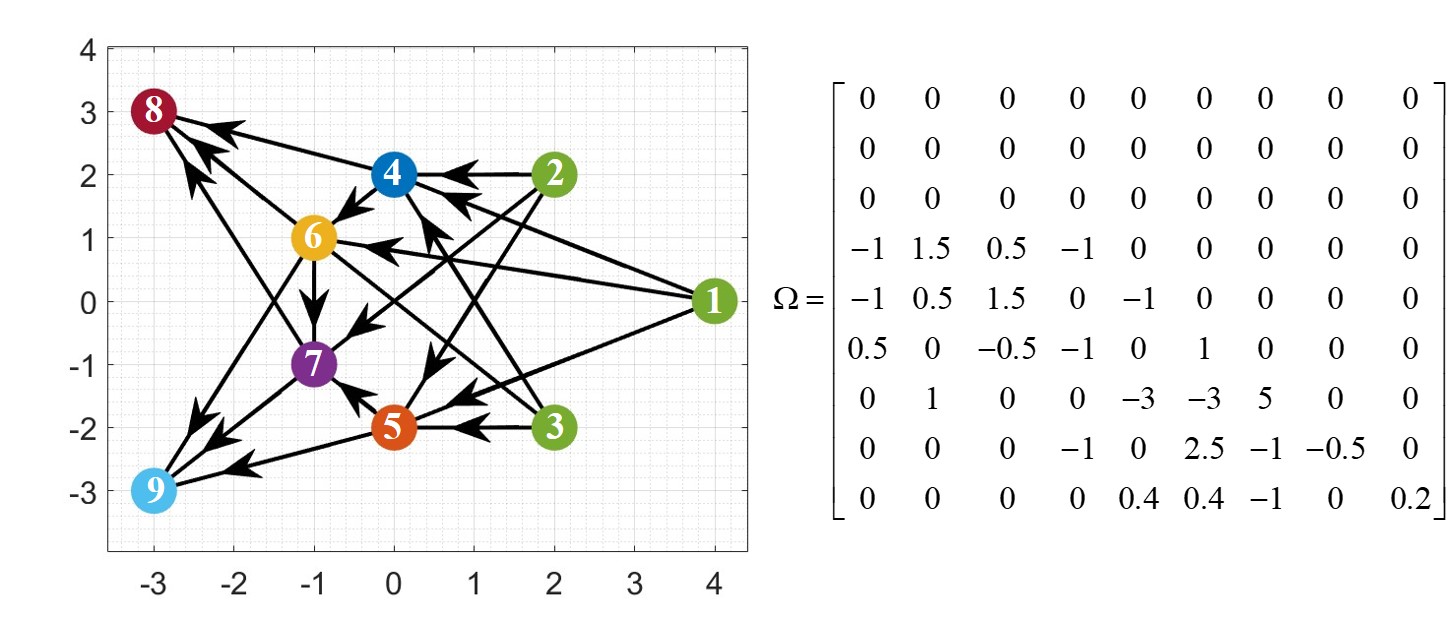}
	\caption{{Nominal framework with a Laplacian matrix.}}
	\label{graphzxt}
\end{figure}

\begin{figure*}[!t]
	\centering
	\includegraphics[width=0.9\linewidth]{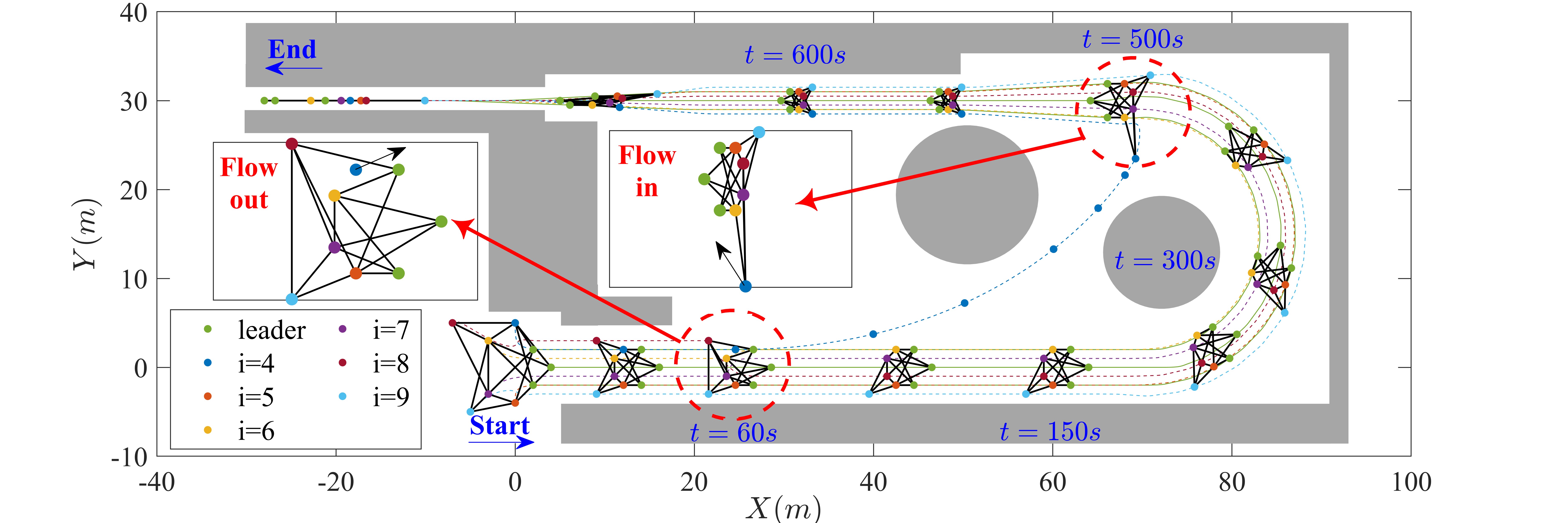}
	\caption{{Formation maneuver trajectories with node flow out and flow in situations.}}
	\label{tra}
\end{figure*}

\begin{figure*}[!t]
	\centering
	\subfloat[]{\includegraphics[width=2.25in]{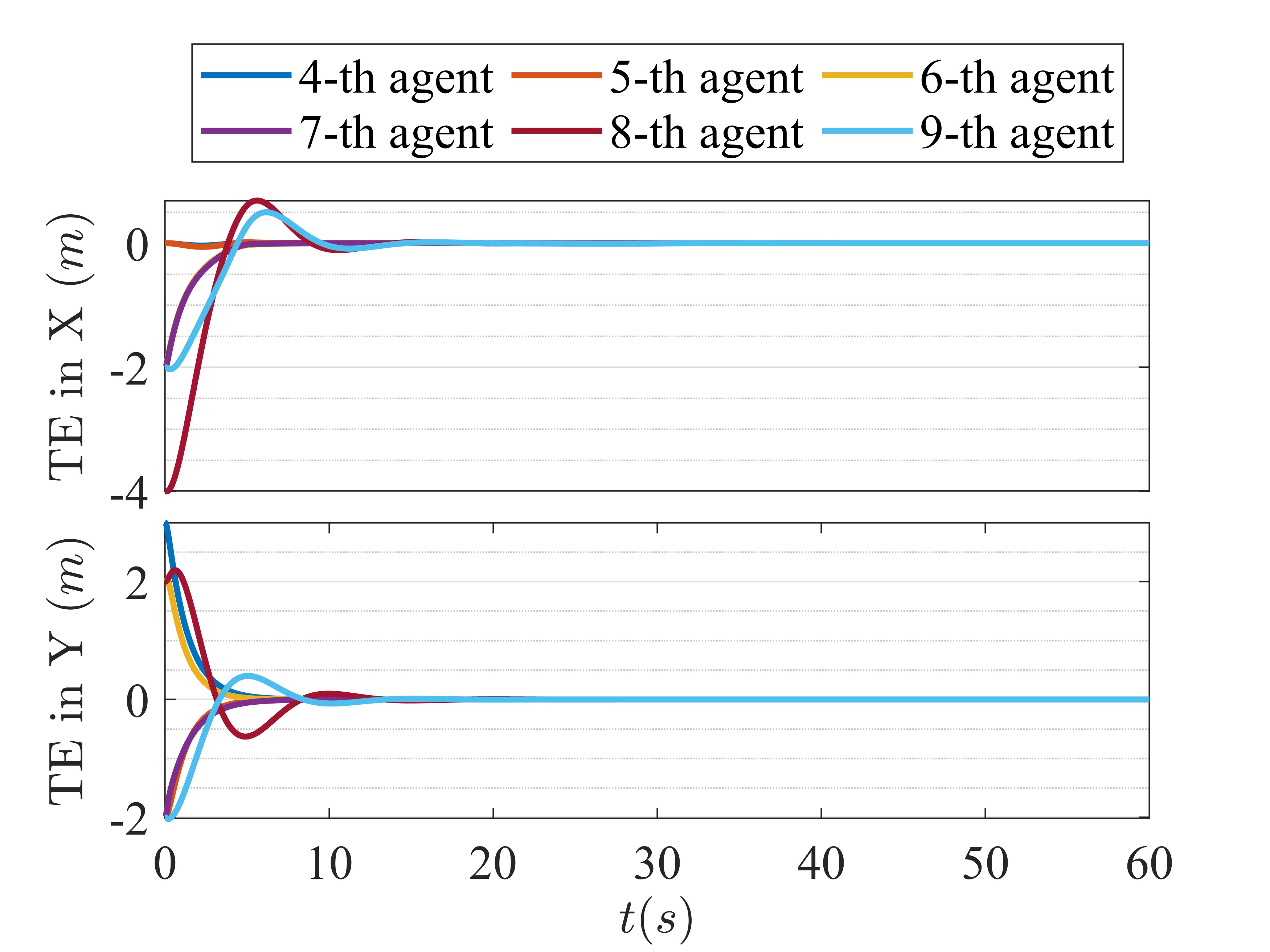}%
		\label{err1}}
	\hfil
	\subfloat[]{\includegraphics[width=2.25in]{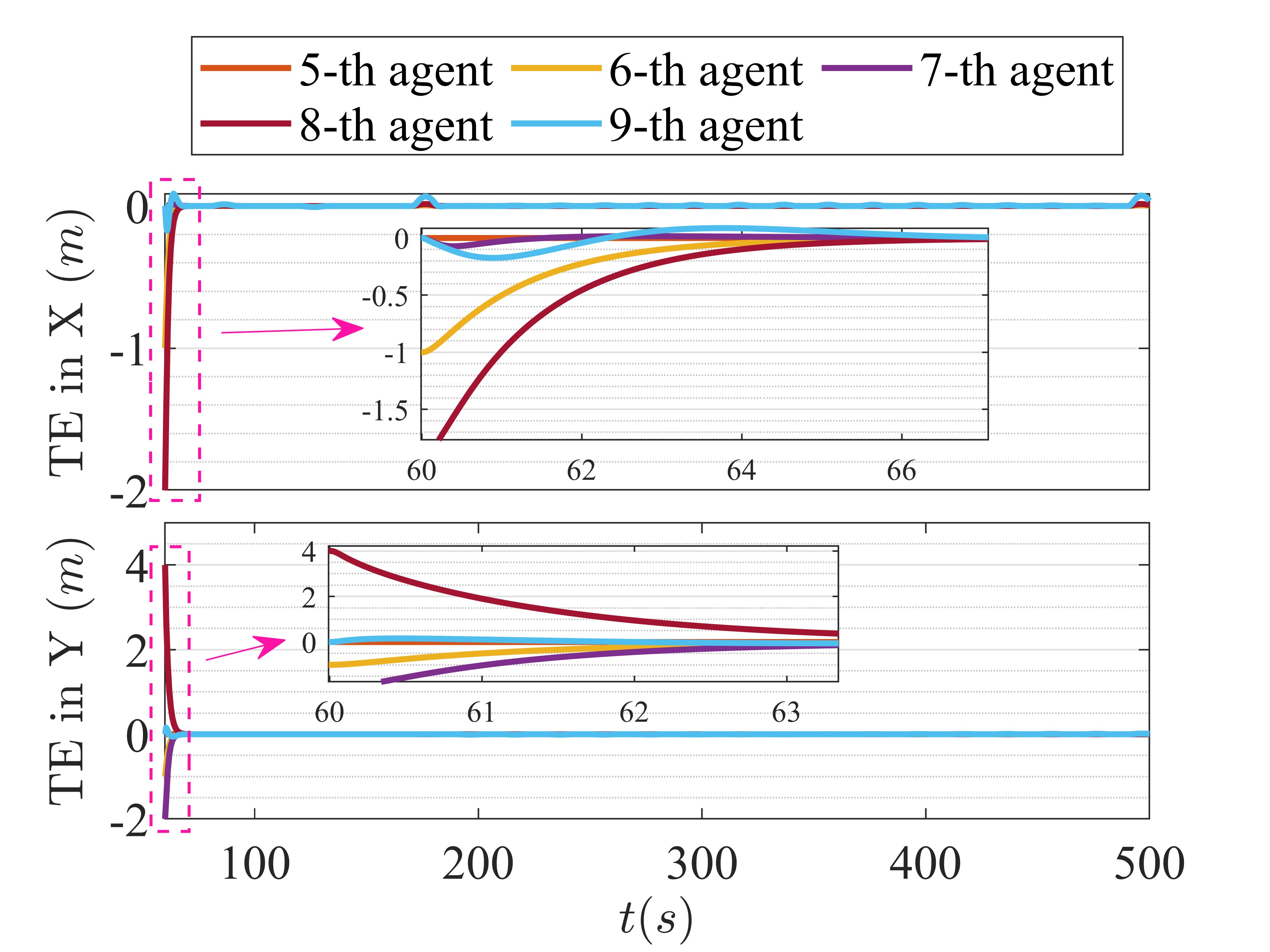}%
		\label{err2}}
	\hfil
	\subfloat[]{\includegraphics[width=2.25in]{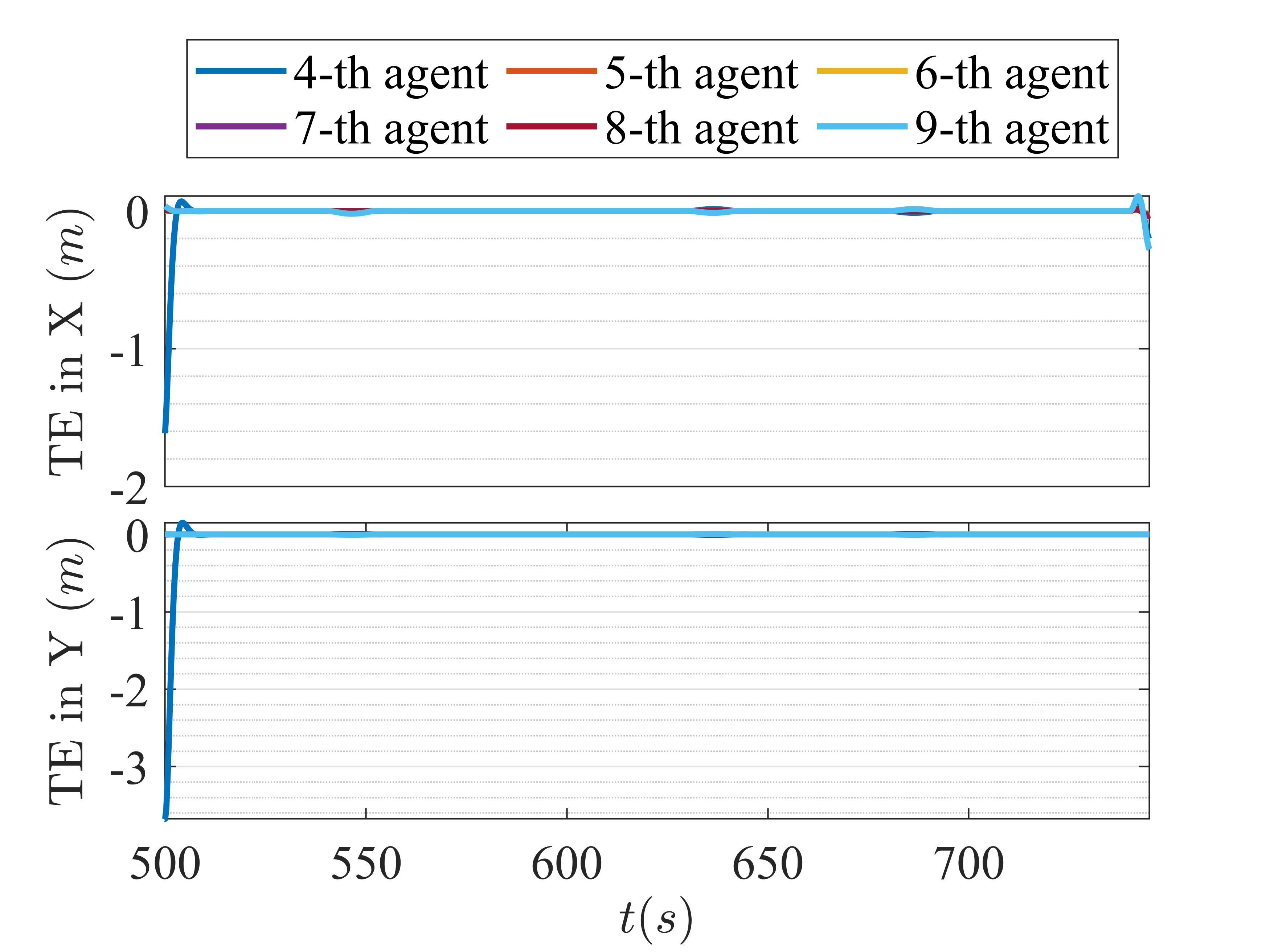}%
		\label{err3}}
	\caption{(a) Formation tracking errors during time interval $[0,60)$. (b) Formation tracking errors during time interval $[60,500)$. (c) Formation tracking errors during time interval $[500,740)$.}
	\label{pic3}
\end{figure*}

\section{Simulation Results}

In this section, a second-order MASs consisting of 3 virtual leaders and 6 controlled followers is adopted to move across an obstacle-laden area. Considering the situations of removing node 4 from the swarm at $t = 60\text{s}$ and adding it again at $t = 500\text{s}$. Nominal framework is shown in Fig. \ref{graphzxt}.

Results are presented in Figs. \ref{tra}-\ref{pic3}. As we can see, MASs successfully maneuvers with translation, rotation, scaling and shearing to avoid collision. When $t \in [0,60)$, MASs form the target formation. At $t = 60\text{s}$, node 4 is removed from the existing framework, by following FOA steps, the nominal framework is reconstructed such that affine localizability of nominal framework is remained. Specifically, inheritance operation is carried out within the following path ${\mathcal{V}_{4}}\rightarrow{\mathcal{V}_{6}}\rightarrow{\mathcal{V}_{7}}\rightarrow{\mathcal{V}_{8}}$. During time interval $[60,500)$, the remaining nodes form a new formation and achieve formation maneuvers successfully. At $t = 500\text{s}$, node 4 is added to the existing framework again, by following FIA steps, a new nominal framework containing node 4 is established. During time interval $[500,740)$, formation maneuvers of the whole MASs is achieved. Thus, Fig. \ref{tra} validate \textit{Theorem 3} and \textit{Theorem 4} of our proposed approach. Formation tracking errors during different time intervals are depicted in Fig. \ref{pic3}, which validate \textit{Theorem 5} of our proposed approach with well convergence property.

\section{Conclusion}
This brief studies affine localizability of MASs from local perspective and its application in affine formation maneuver control. A novel affine localizability structure, called equilibrium unit, is proposed to establish condition and perform construction for MASs. Specifically, layerable directed graph is introduced to replace generic assumption such that the affine localizability condition can be established locally. Furthermore, by applying EUC method, distributed construction manner of affine localizability can be achieved. Different from classical affine formation maneuver control, FIA and FOA equip MASs with self-reconstruction capability when nodes are added to (removed from) the existing nominal framework.

\appendix
\subsection{Proof of Lemma 2}
First, we show that if condition 1) and 2) are satisfied, then, there is $\{h_{j}\in\mathbb{R}_{\ne 0}\}_{j=1}^{M_{d}}$ satisfying \eqref{1}. Then, the displacement matrix is defined as
\begin{equation} \label{weights 1}
	P(z) = \begin{bmatrix}
		z_{1}^{\text{T}} \\
		z_{2}^{\text{T}} \\
		\vdots \\
		z_{M_{d}}^{\text{T}}
	\end{bmatrix}. \tag{A1}
\end{equation}
Then, solving \eqref{1} reduces to finding the kerner space of $(P(z))^{\text{T}}$. Since $\forall R_{j} = R / \{r_{j}\}$, $j = 0,1,\ldots,M_{d}$ is affinely independent and $Z = \{z_{i}\}_{i = 1}^{M_{d}}$ is linearly dependent, then, any $M_{d}-1$ row vectors from $P(z)$ are linearly independent and $\text{Rank}(P(z)) = M_{d} - 1$. Then, we can derive $\text{dim}(\text{Ker}((P(z))^{\text{T}})) = 1$. It means that $\text{Ker}((P(z))^{\text{T}})$ has and only has one non-zero basis $v = \text{col}\{v_{1},v_{2},\ldots,v_{M_{d}}\}\in\mathbb{R}^{M_{d}}$. As a result, the solution to \eqref{1} is $c_{b}v$, where $c_{b}\in\mathbb{R}$ can be chosen arbitrarily. Mathematically, we have $c_{b}(P(z))^{\text{T}}v = 0_{d}$, i.e.,
\begin{equation} \label{weights 2}
	v_{1}z_{1} + v_{2}z_{2} + \ldots + v_{M_{d}}z_{M_{d}} = 0_{d}, \tag{A2}
\end{equation}
To achieve \eqref{weights 2}, when any component of $v$ is zero, the remaining components must also be zero, as any $M_{d} - 1$ row vectors from $P(z)$ are linearly independent. For instant, when $v_{1}=0$, then, $\sum^{M_{d}}_{j=2} v_{j}z_{j} = 0_{d}$ if and only if $v_{2} = v_{3} = \ldots = v_{M_{d}} = 0$, since $z_{2}, z_{3}, \ldots, z_{M_{d}}$ are linear independently. Thus, there is no component of $v$ is zero.

Now, we show that if condition 1) and 2) are satisfied, then, the weight set $\{h_{j}\in\mathbb{R}_{\ne 0}\}_{j=1}^{M_{d}}$ that satisfies \eqref{1} can also satisfy \eqref{2}. If $\{h_{j}\in\mathbb{R}_{\ne 0}\}_{j=1}^{M_{d}}$ satisfies \eqref{1}, then, $\text{col}\{h_{j}\}_{j=1}^{M_{d}} \in \text{Ker}((P(z))^{\text{T}})$. Owing to $\text{Ker}((P(z))^{\text{T}}) \perp \text{Im}(P(z))$, the problem reduces to verifying $1_{M_{d}} \notin \text{Im}(P(z))$. The augmented matrix is defined as
\begin{equation} \label{weights 3}
	P(\bar z) = \begin{bmatrix}
		\bar{z}_{1}^{\text{T}} \\
		\bar{z}_{2}^{\text{T}} \\
		\vdots \\
		\bar{z}_{M_{d}}^{\text{T}}
	\end{bmatrix}, \tag{A3}
\end{equation}
where $\bar{z}_{i} = \text{col}\{z_{i},1\}$. If $1_{M_{d}} \in \text{Im}(R)$, then there is a nonzero $v^{*}\in\mathbb{R}^{d}$ such that
\begin{equation} \label{weights 4}
	v^{*}_{1} [P(z)]_{1:M_{d},1} + v^{*}_{2} [P(z)]_{1:M_{d},2} + \ldots + v^{*}_{d} [P(z)]_{1:M_{d}, d} = 1_{M_{d}}. \tag{A4}
\end{equation}
From \eqref{weights 3}-\eqref{weights 4}, we have $\text{Rank}(P(\bar z)) = M_{d} - 1$, which contradicts the condition 1). Thus, $1_{M_{d}} \notin \text{Im}(R)$ holds. $\blacksquare$

\subsection{Proof of Lemma 3}
It is obvious that $\mathcal{H}_{1}$ is consisted of nodes which have no in-neighbors. Assume that $\mathcal{V}_{j_{1}} \rightarrow \mathcal{V}_{j_{2}} \rightarrow \ldots \rightarrow \mathcal{V}_{j_{M}}$ is the longest path from $\mathcal{H}_{1}$ to an end node. Then, we put $\mathcal{V}_{j_{2}},\mathcal{V}_{j_{3}},\ldots,\mathcal{V}_{j_{M}}$ to $\mathcal{H}_{{2}},\mathcal{H}_{{3}},\ldots,\mathcal{H}_{{M}}$, respectively. For any other nodes, such as $\mathcal{V}_{k}$, let $M_{k}$ be the number of involved nodes corresponding to the longest path from $\mathcal{H}_{1}$ to $\mathcal{V}_{k}$, the we put $\mathcal{V}_{k}$ to $\mathcal{H}_{{M_{k}}}$. By repeating this operation, we can finally uniquely classify all the nodes into hierarchies $\mathcal{H}_{1},\mathcal{H}_{{2}},\ldots,\mathcal{H}_{{M}}$. $\blacksquare$

\subsection{Proof of Theorem 1}
It is obvious that condition 1) in \textit{Definition 3} is satisfied if $\{\chi_{i}\}_{i=1}^{d+1}$ affinely span $\mathbb{R}^{d}$. Then, we further verify that there exist a Laplacian matrix such that $\Omega_{ff}$ is invertible.

Assume that the non-zero weights corresponding to every equilibrium unit are calculated by \eqref{1} and \eqref{2}. Then, we can obtain a Laplacian matrix $\Omega$ as follows
\begin{equation}\label{laplacian matrix1}
	\begin{aligned}
	&\Omega = \\
	&\begin{bmatrix}
		0 & \ldots & 0 & \multicolumn{1}{|c}{0} & \ldots & 0\\
		\vdots & \vdots & \vdots & \multicolumn{1}{|c}{\vdots} & \vdots & \vdots\\
		0 & \ldots & 0 & \multicolumn{1}{|c}{0} & \ldots & 0 \\
		\cline{1-3}
		-w_{N_{l} + 1,1} & \ldots & -w_{N_{l} + 1,N_{l}} & \Sigma^{w}_{N_{l} + 1} & \ldots & -w_{N_{l} + 1,N} \\
		\vdots & \vdots & \vdots & \vdots & \ddots & \vdots\\
		-w_{N,1} & \ldots & -w_{N,N_{l}} & -w_{N,N_{l} + 1} & \ldots & \Sigma^{w}_{N}
	\end{bmatrix}, 
	\end{aligned} \tag{C1}
\end{equation}
where $\Sigma^{w}_{i} = \sum\nolimits_{j \in \mathcal{N}_{i}^{\text{in}}} w_{i,j}$, $i = N_{l} + 1, N_{l} + 2, \ldots, N$. Before verifying the invertibility of $\Omega_{ff}$, we first define the hierarchies of nodes in $\mathcal{G}$. Assume that $\mathcal{H}_{1},\mathcal{H}_{2},\ldots,\mathcal{H}_{M}$ are hierarchies of $\mathcal{G}$, it is obvious that $\mathcal{H}_{1} = \{\mathcal{V}_{1},\mathcal{V}_{2},\ldots,\mathcal{V}_{N_{l}}\}$. Nodes in $\mathcal{H}_{l}$, $l \ge 2$ are denoted from $\mathcal{V}_{^{l}v_{1}}$ to $\mathcal{V}_{^{l}v_{m_{l}}}$. Then, we perform the row and column elementary transformation on \eqref{laplacian matrix1} such that row and column order of \eqref{laplacian matrix1} is coincident with the order in $\mathcal{H}_{1},\mathcal{H}_{2},\ldots,\mathcal{H}_{M}$. Based on this operation, we can obtain that $\Omega_{ff}$ can be transformed into lower triangular $\bar{\Omega}_{ff}$ by elementary transformation. Besides, according to \eqref{2}, the diagonal elements of $\bar{\Omega}_{ff}$ are all non-zero, which implies that $\Omega_{ff}$ is invertible. $\blacksquare$

\subsection{Proof of Theorem 2}
From \textit{step 1} in \textit{Theorem 1}, we have $\{\chi_{i}\}_{i=1}^{d+1}$ affinely span $\mathbb{R}^{d}$, which implies that condition 1) in \textit{Definition 3} is satisfied. Then, we further verify that $\Omega_{ff}$ is invertible in the subsequent part.

Specifically, we can get a Laplacian matrix $\Omega$ in the form of \eqref{laplacian matrix} by performing the steps described in \textit{Theorem 1}.
\begin{equation}\label{laplacian matrix}
	\Omega = 
	\begin{bmatrix}
		0 & \ldots & 0 & \multicolumn{1}{|c}{0} & \ldots & 0\\
		\vdots & \vdots & \vdots & \multicolumn{1}{|c}{\vdots} & \vdots & \vdots\\
		0 & \ldots & 0 & \multicolumn{1}{|c}{0} & \ldots & 0 \\
		\cline{1-3}
		-w_{d+2,1} & \ldots & -w_{d+2,d+1} & \Sigma^{w}_{d+2} & \ldots & 0 \\
		\vdots & \vdots & \vdots & \vdots & \ddots & \vdots\\
		-w_{N,1} & \ldots & -w_{N,d+1} & -w_{N,d+2} & \ldots & \Sigma^{w}_{N}
	\end{bmatrix} , \tag{D1}
\end{equation}
where $\Sigma^{w}_{i} = \sum\nolimits_{j \in \mathcal{N}_{i}^{\text{in}}} w_{i,j}$, $i = d+2, d+3, \ldots, N$. According to \eqref{1} and \eqref{2}, we have $\Sigma^{w}_{i} \ne 0$, $i = d+2, d+3, \ldots, N$, which implies that $\text{Rank}(\Omega_{ff}) = N - (d + 1)$, i.e., $\Omega_{ff}$ is invertible. $\blacksquare$

\subsection{Proof of Theorem 4}
\textit{Case 1: ${\mathcal{V}_{k}}$ is an end node.} Then, the column and row corresponding to ${\mathcal{V}_{k}}$ in the Laplacian matrix $\Omega$ is expressed as follows
\begin{equation}\label{laplacian matrix 2}
	\Omega = 
	\begin{bmatrix}
		 &  &  & 0 &  & &\\
		 &  &  & \vdots &  & &\\
		 &  &  & 0 &  & &\\
		-w_{k,1} & \ldots & -w_{k,(k - 1)} & \Sigma^{w}_{{k}} & 0 & \ldots & 0\\
		&  &  & 0 &  & &\\
		 &  &  & \vdots &  & & \\
		 &  &  & 0 &  & &
	\end{bmatrix} , \tag{E1}
\end{equation}
since ${\mathcal{V}_{k}}$ is the end node, thus, the column elements corresponding to ${^{F}\mathcal{V}_{k}}$ are all zeros except $\Sigma^{w}_{k}$. Then, by implementing the FOA methods, we can get $\Omega^{-}$ by deleting the column and row corresponding to ${\mathcal{V}_{k}}$ described in \eqref{laplacian matrix 2}. It's obvious that leaders' affinely span property is preserved and $\text{Rank}(\Omega_{ff}^{-}) =  N - 1 - (d + 1)$, thus $\Omega_{ff}^{-}$ is invertible. Consequently, the affine localizability of $(\mathcal{G}^{-},\chi^{-})$ is restored.

\textit{Case 2: ${\mathcal{V}_{k}}$ is not an end node.} Consider there exists a path $\mathcal{V}_{k} \rightarrow \mathcal{V}_{j_{1}} \rightarrow \mathcal{V}_{j_{2}} \rightarrow \ldots$, this path allows the information from $\mathcal{V}_{k}$ to be transferred to $\mathcal{V}_{j_{1}}$. Thus, we can replace $\mathcal{V}_{k}$ with $\mathcal{V}_{j_{1}}$ when $\mathcal{V}_{k}$ is removed, and then replacing $\mathcal{V}_{j_{1}}$ with $\mathcal{V}_{j_{2}}$, and continue this process until the end vertex is shifted forward. This process essentially equals to removing an end vertex, and as shown in the analysis above, affine localizability remains unaffected. $\blacksquare$

\bibliographystyle{IEEEtran}
\bibliography{IEEEabrv,Bibliography}

\begin{thebibliography}{10}
\providecommand{\url}[1]{#1}
\csname url@samestyle\endcsname
\providecommand{\newblock}{\relax}
\providecommand{\bibinfo}[2]{#2}
\providecommand{\BIBentrySTDinterwordspacing}{\spaceskip=0pt\relax}
\providecommand{\BIBentryALTinterwordstretchfactor}{4}
\providecommand{\BIBentryALTinterwordspacing}{\spaceskip=\fontdimen2\font plus
\BIBentryALTinterwordstretchfactor\fontdimen3\font minus \fontdimen4\font\relax}
\providecommand{\BIBforeignlanguage}[2]{{%
\expandafter\ifx\csname l@#1\endcsname\relax
\typeout{** WARNING: IEEEtran.bst: No hyphenation pattern has been}%
\typeout{** loaded for the language `#1'. Using the pattern for}%
\typeout{** the default language instead.}%
\else
\language=\csname l@#1\endcsname
\fi
#2}}
\providecommand{\BIBdecl}{\relax}
\BIBdecl

\bibitem{ren2006consensus}
W.~Ren, ``Consensus based formation control strategies for multi-vehicle systems,'' in \emph{2006 American Control Conference}.\hskip 1em plus 0.5em minus 0.4em\relax IEEE, 2006, pp. 6--pp.

\bibitem{cheng2018fully}
B.~Cheng and Z.~Li, ``Fully distributed event-triggered protocols for linear multiagent networks,'' \emph{IEEE Transactions on Automatic Control}, vol.~64, no.~4, pp. 1655--1662, 2018.

\bibitem{pan2023improved}
Y.~Pan, W.~Ji, H.-K. Lam, and L.~Cao, ``An improved predefined-time adaptive neural control approach for nonlinear multiagent systems,'' \emph{IEEE transactions on automation science and engineering}, vol.~21, no.~4, pp. 6311--6320, 2023.

\bibitem{aranda2015coordinate}
M.~Aranda, G.~L{\'o}pez-Nicol{\'a}s, C.~Sag{\"u}{\'e}s, and M.~M. Zavlanos, ``Coordinate-free formation stabilization based on relative position measurements,'' \emph{Automatica}, vol.~57, pp. 11--20, 2015.

\bibitem{zhao2015bearing}
S.~Zhao and D.~Zelazo, ``Bearing rigidity and almost global bearing-only formation stabilization,'' \emph{IEEE Transactions on Automatic Control}, vol.~61, no.~5, pp. 1255--1268, 2015.

\bibitem{chen2020angle}
L.~Chen, M.~Cao, and C.~Li, ``Angle rigidity and its usage to stabilize multiagent formations in 2-d,'' \emph{IEEE Transactions on Automatic Control}, vol.~66, no.~8, pp. 3667--3681, 2020.

\bibitem{lin2015necessary}
Z.~Lin, L.~Wang, Z.~Chen, M.~Fu, and Z.~Han, ``Necessary and sufficient graphical conditions for affine formation control,'' \emph{IEEE Transactions on Automatic Control}, vol.~61, no.~10, pp. 2877--2891, 2015.

\bibitem{zhao2018affine}
S.~Zhao, ``Affine formation maneuver control of multiagent systems,'' \emph{IEEE Transactions on Automatic Control}, vol.~63, no.~12, pp. 4140--4155, 2018.

\bibitem{xu2020affine}
Y.~Xu, S.~Zhao, D.~Luo, and Y.~You, ``Affine formation maneuver control of high-order multi-agent systems over directed networks,'' \emph{Automatica}, vol. 118, p. 109004, 2020.

\bibitem{zhu2022distributed}
C.~Zhu, B.~Huang, Y.~Lu, X.~Li, and Y.~Su, ``Distributed affine formation maneuver control of autonomous surface vehicles with event-triggered data transmission mechanism,'' \emph{IEEE Transactions on Control Systems Technology}, vol.~31, no.~3, pp. 1006--1017, 2022.

\bibitem{zhao2023specified}
Y.~Zhao, K.~Gao, P.~Huang, and G.~Chen, ``Specified-time affine formation maneuver control of multiagent systems over directed networks,'' \emph{IEEE Transactions on Automatic Control}, vol.~69, no.~3, pp. 1936--1943, 2023.

\bibitem{chen2020distributed}
L.~Chen, J.~Mei, C.~Li, and G.~Ma, ``Distributed leader--follower affine formation maneuver control for high-order multiagent systems,'' \emph{IEEE Transactions on Automatic Control}, vol.~65, no.~11, pp. 4941--4948, 2020.

\bibitem{xiao2022framework}
F.~Xiao, Q.~Yang, X.~Zhao, and H.~Fang, ``A framework for optimized topology design and leader selection in affine formation control,'' \emph{IEEE Robotics and Automation Letters}, vol.~7, no.~4, pp. 8627--8634, 2022.

\bibitem{gao2022practical}
K.~Gao, Y.~Liu, Y.~Zhou, Y.~Zhao, and P.~Huang, ``Practical fixed-time affine formation for multi-agent systems with time-based generators,'' \emph{IEEE Transactions on Circuits and Systems II: Express Briefs}, vol.~69, no.~11, pp. 4433--4437, 2022.

\bibitem{li2020layered}
D.~Li, G.~Ma, Y.~Xu, W.~He, and S.~S. Ge, ``Layered affine formation control of networked uncertain systems: A fully distributed approach over directed graphs,'' \emph{IEEE Transactions on Cybernetics}, vol.~51, no.~12, pp. 6119--6130, 2020.

\bibitem{wang2021affine}
J.~Wang, X.~Ding, C.~Wang, L.~Liang, and H.~Hu, ``Affine formation control for multi-agent systems with prescribed convergence time,'' \emph{Journal of the Franklin Institute}, vol. 358, no.~14, pp. 7055--7072, 2021.

\bibitem{lin2021unified}
Y.~Lin, Z.~Lin, Z.~Sun, and B.~D. Anderson, ``A unified approach for finite-time global stabilization of affine, rigid, and translational formation,'' \emph{IEEE Transactions on Automatic Control}, vol.~67, no.~4, pp. 1869--1881, 2021.

\bibitem{alfakih2011bar}
A.~Y. Alfakih, ``On bar frameworks, stress matrices and semidefinite programming,'' \emph{Mathematical programming}, vol. 129, no.~1, pp. 113--128, 2011.

\bibitem{zhang2023self}
Y.~Zhang, S.~O{\u{g}}uz, S.~Wang, E.~Garone, X.~Wang, M.~Dorigo, and M.~K. Heinrich, ``Self-reconfigurable hierarchical frameworks for formation control of robot swarms,'' \emph{IEEE Transactions on Cybernetics}, vol.~54, no.~1, pp. 87--100, 2023.

\bibitem{zhou2024affine}
X.~Zhou, B.~Huang, B.~Zhou, C.~Zhu, H.~Qin, and J.~Miao, ``Affine formation maneuver control for nusvs: An anti-competing interaction solution with random packet losses,'' \emph{IEEE Transactions on Automation Science and Engineering}, 2024.

\end{thebibliography}

\end{document}